\documentclass[12pt,draftcls,onecolumn]{IEEEtran}
\usepackage{cite}
\usepackage[font=small]{caption}
\usepackage{ifpdf}
\usepackage{authblk}
\usepackage{commath}
\usepackage{cuted} 

\usepackage{bigints}%\stripsep=6pt % Po
\newif\ifCLASSOPTIONonecolumn       \CLASSOPTIONonecolumnfalse
\newif\ifCLASSOPTIONtwocolumn       \CLASSOPTIONtwocolumntrue
\ifCLASSINFOpdf
   \usepackage[pdftex]{graphicx}
 \else
    \usepackage[dvips]{graphicx}
  \fi
\usepackage{subcaption}
\usepackage{amsmath}
\usepackage{dsfont}
\usepackage{bbold}
\usepackage{amssymb}
\usepackage{algorithmic}
\usepackage{array}
\usepackage{stix}
\usepackage{stfloats}
\fnbelowfloat
\usepackage{bigints}
\begin{document}
\title{Performance Analysis of Cooperative Communications at Road Intersections Using Stochastic Geometry Tools}%
\author[1 2]{Baha Eddine Youcef~Belmekki}
\author[1]{Abdelkrim ~Hamza}
\author[2]{Beno\^it~Escrig}
\affil[1]{LISIC Laboratory, Electronic and Computer Faculty, USTHB, Algiers, Algeria,}
\affil[ ]{email: $\{$bbelmekki, ahamza$\}$@usthb.dz}
\affil[2]{University of Toulouse, IRIT Laboratory, School
of ENSEEIHT, Institut National Polytechnique de Toulouse, France, e-mail: $\{$bahaeddine.belmekki, benoit.escrig$\}$@enseeiht.fr}
\affil[ ]{}
\setcounter{Maxaffil}{0}
\renewcommand\Affilfont{\small}
\markboth{ }%
{Shell \MakeLowercase{\textit{et al.}}: Bare Demo of IEEEtran.cls for IEEE Journals}
\maketitle

\IEEEpeerreviewmaketitle
\begin{abstract}

Vehicular safety communications (VSCs) provide relevant solutions to avoid congestion and road accidents, more particularly at road intersections since these areas are more prone to accidents. In this context, one of the main impairments that affect the performance of VSCs are interference. In this paper, we study the performance of cooperative transmissions  at intersections in presence of interference for VSCs. We use tools from stochastic geometry, and model interferer vehicles locations as a Poisson point process. First, we calculate the outage probability (OP) for a direct transmission when the received node can be anywhere on the plan. Then, we analyze the OP performance of a cooperative transmission scheme. The analysis takes into account two dimensions: the decoding strategy at the receiver, and the vehicles mobility. We determine the optimal relay position, for different traffic densities and for different vehicles mobility models. We also show that the OP does not improve after the number of infrastructure relays reach a threshold value. Finally, we show that the OP performance of VSCs is higher at intersections than on highways. We validate our analytical results by Monte Carlo simulations. 

\end{abstract}


\section{Introduction}
%% The very first letter is a 2 line initial drop letter followed
%\IEEEPARstart{A}{Motivation}
\subsection{Motivation}

Road traffic safety is a major issue, especially at road intersections. Studies showed that 50\% of all crashes are in junction roads (intersections) including fatal crashes, injury crashes, and property damage crashes \cite{traficsafety}. This makes intersections critical areas not only for vehicles, but also for pedestrians and cyclists. Vehicular communications offer a wide range of applications to avoid potential accidents, or to warn vehicles of an accident happening in their vicinity. Vehicular communications consist of vehicle-to-vehicle (V2V) communications, and vehicle-to-infrastructure (V2I) communications, in which vehicles interact with infrastructures, e.g. road-side units (RSUs), or base stastion (BS)\cite{sam2016vehicle,singh2019multipath,boquet2018adaptive}. The main limiting factor that can jeopardize V2V and V2I communications, and degrade the performance in terms of outage probability are the interference originated from other transmitting vehicles \cite{haenggi2009interference}. Hence, it is crucial to take into account the  interference dependence when designing safety applications and protocols \cite{ganti2009spatial}. In order to deal with interference, cooperative communications have been shown to reduce the outage probability, and increase the throughput \cite{tanbourgi2013cooperative}.

\subsection{Related Works}

\subsubsection{Interference analysis using direct transmissions} Several works in the literature focus on the effect of interference using tools from stochastic geometry. However, few researches focus on the interference dynamics, and how temporal and spatial dependence between interferer nodes affect the performance. Ganti et al \cite{ganti2009spatial}, derivate the conditional outage probability when the interferers are originated from the same set. In \cite{haenggi2009outage}, Haneggi takes into account three types of interference dependence: node distributions, channel fading gains, and channel access schemes. Schilcher et al, in \cite{schilcher2012temporal}, extended Haneggi's work by introducing a wider range of interference dependence. In \cite{tanbourgi2014effect} and \cite{tanbourgi2014dual}, the authors investigate the performance of maximum ratio combining (MRC) considering the effects of interference dynamics. 
\subsubsection{Interference analysis cooperative transmissions} as for cooperative transmissions considering interfernece dependence, Ikki et al derive the outage probability of a relaying scheme using Decode-and-forward protocol \cite{ikki2013regenerative}. The author in \cite{altieri2014outage} derive the outage probability using Decode-and-forward and Compress-and-forward protocols. In \cite{schilcher2013does}, the authors analyze the performance of a cooperative single-hop scheme with the aid of one relay, and a two-hope cooperative scheme with the aid of two relays, considering dependent and independent interference. Tanbourgi et al. derive in \cite{tanbourgi2013cooperative} the outage probability of a cooperative scheme with a single relay, where the destination combines the signal obtained from the source transmission and from either a relay transmission or a second source transmission if the relay does not decode the source message. The authors in \cite{crismani2015cooperative} extended their work in \cite{schilcher2013does} to a cooperative transmission with multiple relays and multiple packet transmissions.
\subsubsection{Interference analysis in vehicular Communications} As far as V2V and V2I communications are concerned, several works investigated the effect of interference in highway scenarios\cite{blaszczyszyn2009performance,blaszczyszyn2013stochastic,blaszczyszyn2012vehicular}. In \cite{farooq2016stochastic}, the authors derive the expressions for the intensity of concurrent transmitters and the packet success probability in multi-lane highways. The outage probability is obtained in \cite{jiang2016information} for Nakagami-$m$ fading and Rayleigh fading channels, and the results are verified with real datasets. The authors in \cite{tassi2017modeling} derivate the outage probability and rate coverage probability of vehicles, when the line of sight path to the base station is obstructed by large vehicles sharing other highway lanes.\\
However, few works studied the effect of interference in vehicular communications at intersections. Steinmetz et al derivate the success probability when the receiving node and the interferer nodes are aligned on the road \cite{steinmetz2015stochastic}.  In \cite{abdulla2016vehicle}, the authors analyze the success probability for finite road segments under several channel conditions. The authors in \cite{abdulla2017fine} evaluate the average and the fine-grained reliability for interference-limited V2V communications with the use of the Meta distribution. In \cite{jeyaraj2017reliability}, the authors analyze the performance of an orthogonal street system which consists of multiple intersections, and show that, in high-reliability regime, the orthogonal street system behaves like a 1-D Poisson network. However, in low-reliability regime, it behaves like a 2-D Poisson network. The authors in \cite{kimura2017theoretical} derive the outage probability of V2V communications at intersections in the presence of interference with a power control strategy.
In \cite{article}, the authors investigate the impact of a line of sight and non line of sight transmissions at intersections considering Nakagami-$m$ fading channels.
In \cite{WCNC,J3,VTC,J4,WiMob}, the authors respectively study the impact of non-orthogonal multiple access \cite{WCNC,J3}, cooperative non-orthogonal multiple access \cite{VTC,J4}, and maximum ratio combining with NOMA at intersections \cite{WiMob}. The authors further extended their work to millimeter wave vehicular networks using NOMA in \cite{NoMa,NoMa2}.

In this line of research, we study the performance of V2V and V2I communications in intersection scenarios in presence of interference. However, all the aforementioned works that deal with intersection scenarios  assume independent interference, which is not a realistic assumption. They also consider that the receiving nodes are on the roads, which is not the case in V2I communications. As the best of the authors knowledge, there are no prior work that consider an intersection scenario with direct transmission when the received nodes can be anywhere on the plan, and a cooperative transmission considering vehicles mobility which result in dependent and independent interference.

\subsection{Contribution}

In this paper, we focus on direct and cooperative transmissions for intersection scenarios in presence of interference. We develop a framework to model a direct transmission and a relayed transmission between vehicles (V2V) and between vehicles and infrastructure (V2I) at intersections using tools from stochastic geometry and point process theory. We derive the outage probability expression for a direct transmission, when the receiving node can be anywhere on the plan. We then derive the outage probability in a relayed transmission considering different decoding strategies, namely selection combining (SC) and MRC. We also consider two mobility models. The first model is the low speed or static vehicles (LSV) model which assumes that the interferer vehicles move slowly or not at all. The second model is the high speed vehicles (HSV) model which assumes that the interferer vehicles move at high speed. The main contributions of our paper are as follows:
\begin{itemize}
\item 
	We develop a tractable analysis to model V2V and V2I communications for direct transmissions and for cooperative transmissions in intersection scenarios, and we show that cooperative transmissions always enhance the outage probability performance compared to direct transmissions. We study two mobility models, and compare their outage probability and throughput performance under different traffic densities. 
\item
	We evaluate the outage probability when the destination uses SC and MRC, and we show that MRC has a better performance over SC only when the relay is close to the source. We also evaluate the outage probability for several relay positions, and we find the optimal relay position for different traffic conditions and vehicle mobility models; and we show that the outage probability does not improve after the number of infrastructure relays reached a threshold value. 
	
\item	
	We derivate a closed form of the outage probability when the interference are dependent and independent given specific channel conditions. We also obtained closed form of Laplace transform expressions where the relay and/or the destination are located anywhere on the plan, for specific channel conditions.
\item	
	We study the outage probability performance of cooperative transmissions in highways and intersection scenarios. We show that, as the vehicles move closer to intersections, the outage probability increases compared to highway scenarios. However, as the vehicles move away from intersections, highway and intersection scenarios exhibit the same performance, which confirms the statement that intersections are critical areas and more prone to incidents. Finally, we show that, depending on the environment, the interference dependence behaves differently. For instance, suburban intersections have a lower interference dependence compared to urban intersections. This is obtained by studying the impact of the path loss exponent on the outage probability performance.

\end{itemize}

\subsection{Organization}

The rest of this paper is organized as follows. Section II presents our system model. In Section III, the outage behavior is carried out for several scenarios. The Laplace transform expressions are derived in Section IV. Simulations and discussions can be found in Section V. Finally, we conclude this paper in Section VI.

Notation: We denote $\Vert\cdot\Vert$ the Euclidean norm. $\mathbb{P}(A)$ denotes the probability of the event  \textit{A}. $\mathcal{L}_I(\cdot)$ and $\mathbb{E}_I[\cdot]$ denote the Laplace transform and the expectation with respect to the random variable $I$, respectively. $\mathds{1} \lbrace \cdot \rbrace$ denotes the indicator function.

\section{System Model}
We consider a cooperative transmission between a source node $S$ and a destination node $D$, with the help of a relay node $R$. For the sake of convenience, we use $S$, $D$ and $R$ to denote both the nodes and their locations. 

We consider an intersection case with two perpendicular roads, the horizontal road denoted by $X$ and the vertical road denoted by $Y$.
As we consider both V2V and V2I communications\footnote{The Doppler shift and time-varying effect of V2V and V2I channel is beyond the scope of this paper}, any node of the triplet $\lbrace{S, R, D}\rbrace$ can be either on the road (as a vehicle) or outside the road (as part of the communication infrastructure). As shown in Fig.1, the distance between the relay and the intersection (resp. between the destination and the intersection) is denoted by $r$  (resp. $d$), the intersection is the point when the road $X$ and $Y$ intersect, i.e., the point (0,0).

We also consider a set of interfering vehicles that are located on the roads.
We assume that the set of interfering vehicles on axis $X$, denoted by $\Phi_{X}$ (resp. on axis $Y$, denoted by $\Phi_{Y}$) are distributed according to a one-Dimensional Homogeneous Poisson Point Process (1D-HPPP) denoted by $\Phi_{X}\sim$1D-HPPP$(\lambda_{X},x)$ $\big($resp.$\Phi_{Y}$ $\sim$1D-HPPP$(\lambda_{Y},y)$ $\big)$ over the space $\mathcal{B}$, where $x$ and $\lambda_{X}$ (resp. $y$ and $\lambda_{Y}$) are the position of the interferer vehicles and their intensity on the $X$ road (resp. $Y$ road). We denote by $x$ and $y$, both the interferers and their locations. Finally, we consider that the 1D-HPPP can be on infinite road segments, i.e., $\mathcal{B}=\{x \in \mathbb{R},y \in \mathbb{R}\}$, or on a finite road segments, i.e.,  $\mathcal{B}=\{x,y \in \mathbb{R}| \abs{x}<Z,\abs{y}<Z\}$.

The nodes use a slotted ALOHA protocol. The time axis is divided into time-slots and each transmitting node can access the medium at each time-slot with probability $p\in[0,1]$, which performs an independent $p-$thinning of the original point process. 

We use a Decode--and--Forward (DF) transmission scheme \cite{altieri2014outage}, i.e., the node $R$ decodes and re-encodes the message then forwards it. 
We consider a half-duplex transmission during which a transmission occurs during two phases. The duration of each phase is one time--slot. In the first phase, the source broadcasts the message to the relay and destination $(S
\longrightarrow R$ and $S \longrightarrow D)$.

In the second phase, we consider two decoding schemes. For the first one, named SC, the relay transmits to the destination $(R \longrightarrow D)$. For the second one, named MRC, the relay transmits to the destination and the destination adds the power received from the relay and the power received from the source in the first time slot ($R \longrightarrow D$ and $S \longrightarrow D$).
\begin{figure*}[t!]
    \centering
    \begin{subfigure}[b]{0.5\textwidth}
        \centering
        \includegraphics[height=5cm,width=6cm]{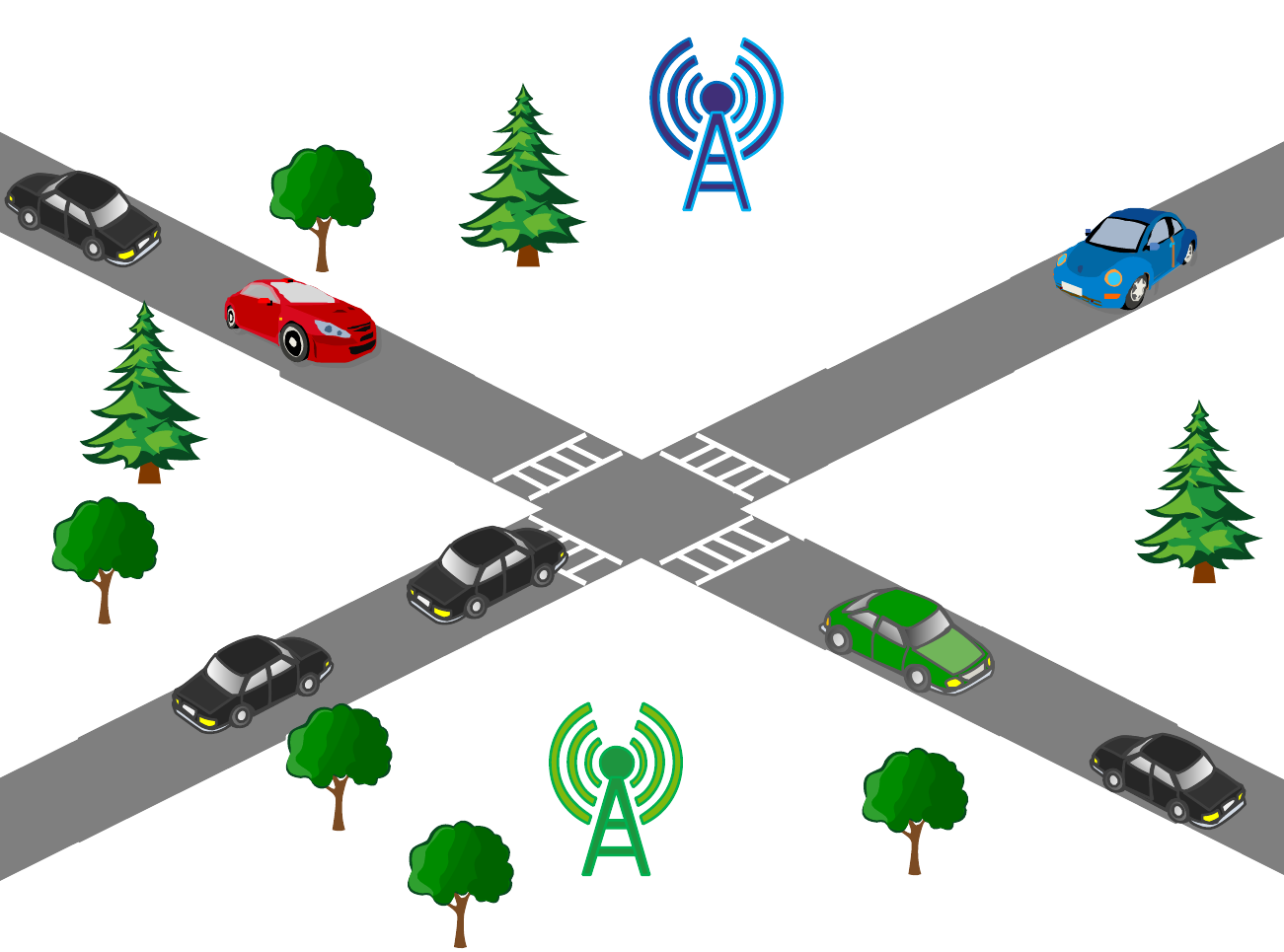}
        \caption{}
    \end{subfigure}%
    ~ 
    \begin{subfigure}[b]{0.5\textwidth}
        \centering
        \includegraphics[height=5cm,width=6cm]{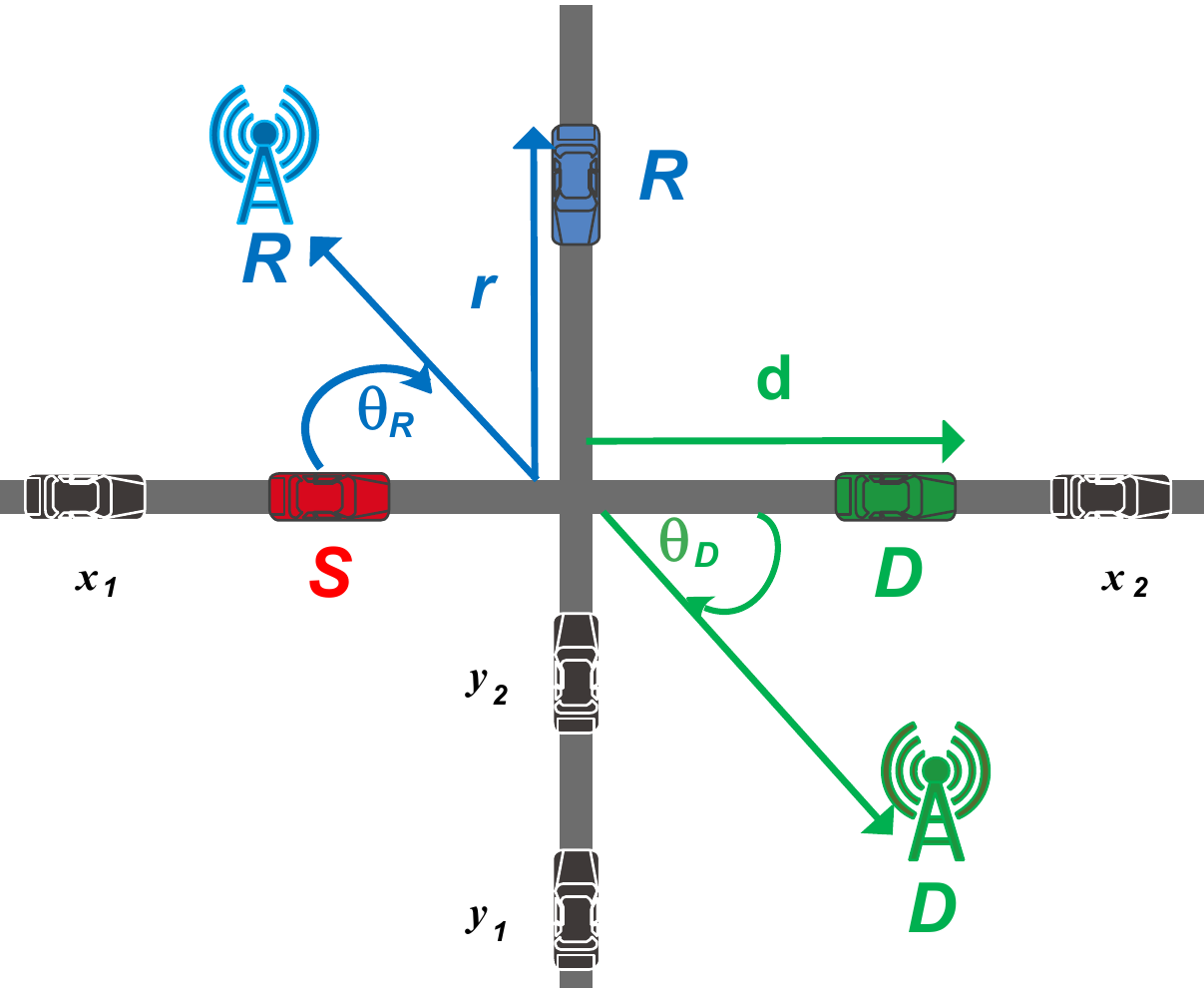}
        \caption{}
    \end{subfigure}
    \caption{System model for vehicular communications involving
a relay $R$ and a destination $D$. The nodes $R$ and $D$ can be vehicles or as part
of the communication infrastructures}
\end{figure*}
The transmission between any pair of two nodes $a$ and $b$ experiences a path loss $l_{ab}=(A r_{ab})^{-\alpha}$, where $A$ is a constant depending on the antenna characteristics, 
$ r_{ab}$ is the Euclidean distance between the node $a$ and $ b$, i.e., $ r_{ab}=\Vert a- b\Vert$, and $\alpha$ is the path loss exponent. All nodes transmit with a constant power $P$, $h_{ab}$ is the fading coefficient between node $a$ and $b$, and is modeled as $\mathcal{CN}(0,1)$ (Rayleigh fading)\cite{cheng2007mobile}. The power fading coefficient between the node $a$ and $b$, $|h_{ab}|^2$, follows an exponential distribution with unit mean. We also consider a Gaussian noise (AWGN) with zero mean and variance $\sigma^2$.
We define the aggregate interference as 
\begin{eqnarray}\label{eq1} 
I_{X_{R}}=\sum_{x\in \Phi_{X_{R}}}P\vert h_{Rx}\vert^{2}l_{Rx}\\ 
I_{Y_{R}}=\sum_{y\in \Phi_{Y_{R}}}P\vert h_{Ry}\vert^{2}l_{Ry} \\
I_{X_{D}}=\sum_{x\in \Phi_{X_{D}}}P\vert h_{Dx}\vert^{2}l_{Dx}\\
I_{Y_{D}}=\sum_{y\in \Phi_{Y_{D}}}P\vert h_{Dy}\vert^{2}l_{Dy},
\end{eqnarray}
where $I_{X_{R}} $ is the aggregate interference from the X road at the relay, $I_{Y_{R}}$ is the aggregate interference from the Y road at the relay, $I_{X_{D}}$ is the aggregate interference from the X road at the destination, $ I_{Y_{D}}$ is the aggregate interference from the Y road at the destination, $\Phi_{X_{R}}$ is the set of the interferers from the road X at the relay, $\Phi_{Y_{R}}$ is the set of the interferers from the road Y at the relay, $\Phi_{X_{D}}$  is the set of the interferers from the road X at the destination, and $\Phi_{Y_{D}}$ is the set of the interferers from the road Y at the destination.

In this work, we consider two mobility models. In the first model, which we referred to as the low speed or static vehicles (LSV), we assume that interferer vehicles do not move or move slowly, that is, their positions remain the same during the two time slots of the transmission. Thus the vehicles that interfere at the relay and at the destination are originated from the same set, i.e., $\Phi_{X_{R}} =\Phi_{X_{D}}=\Phi_{X},\nonumber$ and $\Phi_{Y_{R}}= \Phi_{Y_{D}}=\Phi_{Y}\nonumber$.

In the second model, which we referred to as the high speed vehicles (HSV), we assume that vehicles move at a high speed, that is, their positions change every time slot. Thus, the vehicles that interfere at the relay during the first time slot are not the same as the one that interfere at the destination during the second time slot. This can be modeled by independent realizations of the Poisson point process $\Phi$ for each time slot, i.e., 
$\Phi_{X_{R}} \cap \Phi_{X_{D}}=\varnothing,\nonumber$ and $\Phi_{Y_{R}} \cap \Phi_{Y_{D}}=\varnothing \nonumber$. \\
The HSV model captures the scenario in which the vehicles are highly mobile so that their locations during one time slot do not provide information about their locations during any other time slot.\\
%\begin{flushright}
%\begin{figure}
%\includegraphics[height=10cm,width=12cm]{Fig-a.pdf} 
%%\begin{flushleft}
%\caption{Performance comparison between I}
%%\end{flushleft}
%\label{fig:1}  
%%\vspace{-0.5cm}     % Give a unique label
%\end{figure}
%\end{flushright}
%\begin{figure}
%\includegraphics[height=10cm,width=12cm]{Fig-b.pdf} 
%%\includegraphics[height=7.5cm,width=8cm]{Fig2.pdf} 
%%\begin{flushleft}
%\caption{Performance comparison between 2}
%%\end{flushleft}
%\label{fig:2}  
%%\vspace{-0.5cm}     % Give a unique label
%\end{figure}
\section{Outage Computation}
\subsection{Condition for outage:}
We define the outage events related to the DF protocol that uses a half-duplex transmission. We first define the outage event related to the direct link $S-D$. Then, we define the outage events related to the relayed links $S-R$ and $R-D$ when using SC and MRC\footnote{In this paper, we do not consider the bandwidth in computing the achievable rate.}. 

\subsubsection{Direct link}
We define the rate related to the direct link $S-D$, denoted $\mathcal{R}_{SD}$, as
\begin{equation}\label{eq2} 
\mathcal{R}_{SD}=\dfrac{1}{2}\log_{2}\Bigg(1+\dfrac{P\vert h_{SD}\vert^{2}l_{SD}}{\sigma^{2}+I_{X_{D}}+I_{Y_{D}}}\Bigg).  
\end{equation}
We define the outage event on the direct link $S-D$, denoted $\textit{O}_{SD}$, as 
\begin{equation}\label{eq3} 
 \textit{O}_{SD} \triangleq  \left[\mathcal{R}_{SD} <\mathcal{R} \right], 
\end{equation}
where $\mathcal{R}$ is the target rate.
 
\subsubsection{Relayed link}
We define the rates related to the relayed link $S-R$ and $R-D$, denoted respectively $\mathcal{R}_{SR}$ and $\mathcal{R}_{RD}$, as

\begin{equation}\label{eq4} 
\mathcal{R}_{SR}=\dfrac{1}{2}\log_{2}\Bigg(1+\dfrac{P\vert h_{SR}\vert^{2}l_{SR}}{\sigma^{2}+I_{X_{R}}+I_{Y_{R}}}\Bigg),  
\end{equation}
and
\begin{equation}\label{eq5} 
\mathcal{R}_{RD}=\dfrac{1}{2}\log_{2}\Bigg(1+\dfrac{P\vert h_{RD}\vert^{2}l_{RD}}{\sigma^{2}+I_{X_{D}}+I_{Y_{D}}}\Bigg).  
\end{equation}
We now define the outage events on the relayed link $S-R$ and $R-D$ as
\begin{equation}\label{eq6} 
\textit{O}_{SR} \triangleq  \left[\mathcal{R}_{SR} <\mathcal{R} \right] 
\end{equation}
\begin{equation}\label{eq7} 
\textit{O}_{RD} \triangleq  \left[\mathcal{R}_{RD} <\mathcal{R} \right]. 
\end{equation}

Notice that the outage can also be expressed in terms of signal-to-interference-plus-noise ratio (SINR), that is, an outage event occurs when the SINR is lower than a decoding threshold $\Theta$, which can be expressed by
\begin{equation} 
{\mbox{SINR}}_{ab}< \Theta, \nonumber
\end{equation}
where ${\mbox{SINR}}_{ab}=\dfrac{P\vert h_{ab}\vert^{2}l_{ab}}{\sigma^{2}+I_{X_{b}}+I_{Y_{b}}}$, $\Theta=2^{2\mathcal{R}}-1$, and $a$ and $b$ are the transmitting and the receiving node, respectively.

Since we use SC and MRC, we have two expressions of the outage event as in \cite{altieri2014outage}. Therefore, we express the outage $O_{\textbf{(SC)}}$  for SC as
\begin{equation}\label{eq8} 
\textit{O}_{\textbf{(SC)}}=[\textit{O}_{SR} \cap \textit{O}_{SD}] \cup [ O_{SR}^{C} \cap \textit{O}_{RD}],
\end{equation}

where $ O_{SR}^{C}\triangleq  \left[\mathcal{R}_{SR}> \mathcal{R} \right]  $ 
is the event that the relay successfully decodes the source message.
We express the outage $\textit{O}_{\textbf{(MRC)}}$ for MRC as
%is the event that the relay successfully decoded the source's message.\\
%And the DF outage for the second scheme as:
\begin{equation}\label{eq9} 
\textit{O}_{\textbf{(MRC)}}=\left[\textit{O}_{SR} \cap \textit{O}_{SD} \right] \cup [ O_{SR}^{C} \cap \textit{O}_{SRD}],
\end{equation}

where $ O_{SRD}\triangleq  \left[\mathcal{R}_{SRD} < \mathcal{R} \right]  $
is the outage event at the destination when the relay and the source transmit simultaneously, and $\mathcal{R}_{SRD}$ is the data rate when using MRC. The rate $\mathcal{R}_{SRD}$ is expressed as
\begin{equation}\label{eq10} 
\mathcal{R}_{SRD}=\dfrac{1}{2}\log_{2}\Bigg(1+\dfrac{P\vert h_{SD}\vert^{2}l_{SD}+P\vert h_{RD}\vert^{2}l_{RD}}{\sigma^{2}+I_{X_{D}}+I_{Y_{D}}}\Bigg). 
\end{equation}
\subsection{Outage behaviour:}
In this section, we calculate the outage probability for the direct transmission and the relayed transmission.
\subsubsection{Direct transmission} 
The outage probability for a direct transmission has been derived in \cite{steinmetz2015stochastic} and is given by 
\begin{equation}\label{eq11} 
\mathbb{P}(\textit{O})= 1-N_{SD}\mathcal{L}_{I_{X_D} }(K_{SD})\mathcal{L}_{I_{Y_D} }(K_{SD}),
\end{equation}
where $K_{ab}=\dfrac{\Theta}{P l_{ab}}$  and  $N_{ab}=\exp(-K_{ab}\sigma^2 )$.
\subsubsection{Relayed transmission} 
The outage probability for the DF protocol using a half-duplex transmission when the destination uses SC, $\mathbb{P}(\textit{O}_{\textbf{(SC)}})$ is
\begin{equation}\label{eq12} 
\mathbb{P}(\textit{O}_{\textbf{(SC)}})=\mathbb{P}(\textit{O}_{SR} \cap \textit{O}_{SD} ) + \mathbb{P}(O_{SR}^C\cap \textit{O}_{RD} ),
\end{equation}
and the outage probability when the destination uses MRC, $\mathbb{P}(\textit{O}_{\textbf{(MRC)}})$ is
%and the outage probability when the source transmits during two phases is
\begin{equation}\label{eq13} 
\mathbb{P}(\textit{O}_{\textbf{(MRC)}})=\mathbb{P}(\textit{O}_{SR} \cap \textit{O}_{SD} ) + \mathbb{P}(O_{SR}^{C}\cap \textit{O}_{SRD} ).
\end{equation}

Now, we calculate each probability in equations (\ref{eq12}) and (\ref{eq13}). First, we calculate the probability $ \mathbb{P}(\textit{O}_{SR} \cap \textit{O}_{SD}) $. This probability is related to the outage during the first phase, its expression does not change whether the destination uses SC or MRC. The expression of $ \mathbb{P}(\textit{O}_{SR} \cap \textit{O}_{SD}) $ is given by
\begin{multline}\label{eq14} 
\mathbb{P}(\textit{O}_{SR} \cap \textit{O}_{SD} )=1-\mathbb{P}(\textit{O}^C_{SR} \cup \textit{O}^C_{SD} )= 1- \big[\mathbb{P}(\textit{O}^C_{SR})+\mathbb{P}(\textit{O}^C_{SD})-\mathbb{P}(\textit{O}^C_{SR} \cap \textit{O}^C_{SD} )\big]\\
= 1- \mathbb{P}(\textit{O}^C_{SR})-\mathbb{P}(\textit{O}^C_{SD})+\mathbb{P}(\textit{O}^C_{SR} \cap \textit{O}^C_{SD}) ,
\end{multline}
where $\mathbb{P}(\textit{O}^{C}_{SD})$ and $\mathbb{P}(\textit{O}^{C}_{SR})$ are given respectively by 
\begin{equation}\label{eq15} 
\mathbb{P}(\textit{O}^{C}_{SD})=N_{SD}\mathcal{L}_{I_{X_D} }(K_{SD})\mathcal{L}_{I_{Y_D} }(K_{SD}),
 \end{equation}
 and
\begin{equation}\label{eq16} 
\mathbb{P}(\textit{O}^{C}_{SR})=N_{SR}\mathcal{L}_{I_{X_R} }(K_{SR})\mathcal{L}_{I_{Y_R} }(K_{SR}).
 \end{equation}
\textbf{Lemma 1.}
\textit{The outage probability $\mathbb{P}(\textit{O}^{C}_{SR} \cap \textit{O}^{C}_{SD}  )$ is given by}
\begin{multline}\label{eq17} 
\mathbb{P}(\textit{O}^{C}_{SR} \cap \textit{O}^{C}_{SD})=N_{SR}N_{SD}  
\mathbb{E}_{I_X}\big[\exp(-K_{SR}I_{X_R}-K_{SD}I_{X_D})\big] \\
\mathbb{E}_{I_Y}\big[\exp(-K_{SR}I_{Y_R}-K_{SD}I_{Y_D})\big].
\end{multline}
\textit{Proof}: See \ref{appA}.\hfill $ \blacksquare $\\
\textbf{Lemma 2.}
\textit{The outage probability, when using SC, denoted $\mathbb{P}(\textit{O}^{C}_{SR} \cap \textit{O}_{RD}  )$, is expressed as}
\begin{multline}\label{eq18} 
\mathbb{P}(\textit{O}^{C}_{SR} \cap \textit{O}_{RD})=N_{SR}\mathbb{E}_{I_X}\big[\exp(-K_{SR}I_{X_R})\big]  
\mathbb{E}_{I_Y}\big[\exp(-K_{SR}I_{Y_R})\big]\\
-N_{SR}N_{RD}  
\mathbb{E}_{I_X}\big[\exp(-K_{SR}I_{X_R}-K_{RD}I_{X_D})\big] 
\mathbb{E}_{I_Y}\big[\exp(-K_{SR}I_{Y_R}-K_{RD}I_{Y_D})\big].
\end{multline}
\textit{Proof}: See \ref{appB}\hfill $ \blacksquare $

%The outage probability, when the source transmits during two phases, denoted  $ \mathbb{P}(\textit{O}^{C}_{SR} \cap \textit{O}^{C}_{SRD}) $ is given by

\textbf{Lemma 3.}
\textit{The outage probability, when using MRC, denoted $ \mathbb{P}(\textit{O}^{C}_{SR} \cap \textit{O}_{SRD}) $ is given by}
\begin{multline}\label{eq19} 
\mathbb{P}(\textit{O}^{C}_{SR} \cap \textit{O}_{SRD})=N_{SR}\mathbb{E}_{I_X}\big[\exp(-K_{SR}I_{X_R})\big]  
\mathbb{E}_{I_Y}\big[\exp(-K_{SR}I_{Y_R})\big]\\
-\dfrac{N_{SR}N_{RD}l_{RD}}{l_{RD}-l_{SD}} 
\mathbb{E}_{I_X}\big[\exp(-K_{SR}I_{X_R}-K_{RD}I_{X_D})\big] 
\mathbb{E}_{I_Y}\big[\exp(-K_{SR}I_{Y_R}-K_{RD}I_{Y_D})\big] \\
+ \dfrac{N_{SR}N_{SD}l_{SD}}{l_{RD}-l_{SD}} 
\mathbb{E}_{I_X}\big[\exp(-K_{SR}I_{X_R}-K_{SD}I_{X_D})\big]\mathbb{E}_{I_Y}\big[\exp(-K_{SR}I_{Y_R}-K_{SD}I_{Y_D})\big].
\end{multline}
\textit{Proof}: See \ref{appC}. \hfill $ \blacksquare $

Note that the outage probability expressions derived before are expressed as a function of the expectation with respect to $ I_X $ and $I_Y  $. The expression of the expectation changes depending on the vehicles mobility models, i.e., the HSV and the LSV models.\\
\textbf{Theorem 1.}
\textit{The outage probability for the HSV model using SC and MRC is given by (\ref{eq12}) and (\ref{eq13}), where $\mathbb{P}(\textit{O}^{C}_{SR} \cap \textit{O}^{C}_{RD})$ is given by}
\begin{multline}\label{eq20} 
\mathbb{P}(\textit{O}_{SR} \cap \textit{O}_{SD})= 1-
N_{SD}\mathcal{L}_{I_{X_D} }(K_{SD})\mathcal{L}_{I_{Y_D} }(K_{SD})
-N_{SR}\mathcal{L}_{I_{X_R} }(K_{SR})\mathcal{L}_{I_{Y_R} }(K_{SR})\\
 +N_{SR}N_{SD}\mathcal{L}_{I_{X_R} }(K_{SR})\mathcal{L}_{I_{Y_R}}(K_{SR})
 \mathcal{L}_{I_{X_D} }(K_{SD})\mathcal{L}_{I_{Y_D} }(K_{SD}), 
 \end{multline}
\textit{and the probability $\mathbb{P}(\textit{O}^{C}_{SR} \cap \textit{O}_{RD})$ and $\mathbb{P}(\textit{O}^{C}_{SR} \cap \textit{O}_{SRD})$ in (\ref{eq12}) and (\ref{eq13}) are respectively expressed by}
 \begin{multline}\label{eq21} 
\mathbb{P}(\textit{O}^{C}_{SR} \cap \textit{O}_{RD})=N_{SR}\mathcal{L}_{I_{X_R} }(K_{SR})\mathcal{L}_{I_{Y_R} }(K_{SR})\\
 -N_{SR}N_{RD}\mathcal{L}_{I_{X_R} }(K_{SR})\mathcal{L}_{I_{Y_R}}(K_{SR})
 \mathcal{L}_{I_{X_D} }(K_{RD})\mathcal{L}_{I_{Y_D} }(K_{RD}), 
 \end{multline}
 \textit{and}
 \begin{multline}\label{eq22} 
\mathbb{P}(\textit{O}^{C}_{SR} \cap \textit{O}_{SRD})
=N_{SR}\mathcal{L}_{I_{X_R} }(K_{SR})\mathcal{L}_{I_{Y_R} }(K_{SR})\\
 -\dfrac{N_{SR}N_{RD}l_{RD}}{l_{RD}-l_{SD}}\mathcal{L}_{I_{X_R} }(K_{SR})\mathcal{L}_{I_{Y_R} }(K_{SR})\mathcal{L}_{I_{X_D} }(K_{RD})
 \mathcal{L}_{I_{Y_D}}(K_{RD})
 \\
 +\dfrac{N_{SR}N_{SD}l_{SD}}{l_{RD}-l_{SD}}\mathcal{L}_{I_{X_R} }(K_{SR})\mathcal{L}_{I_{Y_R} }(K_{SR})\mathcal{L}_{I_{X_D} }(K_{SD})
 \mathcal{L}_{I_{Y_D}}(K_{SD})
 .
 \end{multline}

 \textit{Proof}:  See \ref{appD}.\hfill $ \blacksquare $\\
\textbf{ Theorem 2.} \textit{The outage probability for the LSV model using SC and MRC is given by (\ref{eq12}) and (\ref{eq13}),  $\mathbb{P}(\textit{O}^{C}_{SR} \cap \textit{O}^{C}_{RD})$ is given by}

 \begin{equation}\label{eq233} 
\mathbb{P}(\textit{O}^{C}_{SR} \cap \textit{O}^{C}_{SD})= 
 N_{SR}N_{SD} \mathcal{L}_{I_{X_R},I_{X_D}}(K_{SR},K_{SD})\mathcal{L}_{I_{Y_R},I_{Y_D}}
 (K_{SR},K_{SD}).
 \end{equation}

 \textit{and the probability $ \mathbb{P}(\textit{O}^{C}_{SR} \cap \textit{O}_{RD}) $ and $ \mathbb{P}(\textit{O}^{C}_{SR} \cap \textit{O}_{SRD}) $ in (\ref{eq12}) and (\ref{eq13}) are respectively given by}
\begin{multline}\label{eq20} 
\mathbb{P}(\textit{O}_{SR} \cap \textit{O}_{SD})= 1-
N_{SD}\mathcal{L}_{I_{X_D} }(K_{SD})\mathcal{L}_{I_{Y_D} }(K_{SD})
-N_{SR}\mathcal{L}_{I_{X_R} }(K_{SR})\mathcal{L}_{I_{Y_R} }(K_{SR})\\
 +N_{SR}N_{SD}\mathcal{L}_{I_{X_R} }(K_{SR})\mathcal{L}_{I_{X_D} }(K_{SD})
 \mathcal{L}_{I_{X_R},I_{X_D}} (K_{SR},K_{SD})\mathcal{L}_{I_{Y_R},I_{Y_D}} (K_{SR},K_{SD}),
 \end{multline}
 \textit{and}
 \begin{multline}\label{eq25} 
\mathbb{P}(\textit{O}^{C}_{SR} \cap \textit{O}_{SRD})= N_{SR}\mathcal{L}_{I_{X_R} }(K_{SR})\mathcal{L}_{I_{Y_R} }(K_{SR})\\ -\dfrac{ N_{SR} N_{RD}l_{RD}}{ l_{RD}- l_{SD}}
\mathcal{L}_{I_{X_R},I_{X_D}}(K_{SR},K_{RD})\mathcal{L}_{I_{Y_R},I_{Y_D}}(K_{SR},K_{RD})\\
+\dfrac{ N_{SR} N_{SD}l_{SD}}{ l_{RD}- l_{SD}} \mathcal{L}_{I_{X_R},I_{X_D}} (K_{SR},K_{SD}) \mathcal{L}_{I_{Y_R},I_{Y_D}} (K_{SR},K_{SD}),
  \end{multline}
\textit{where}
\begin{equation}\label{eq26.1}
  \mathcal{L}_{I_{X_R},I_{X_D}}(s,b)=\mathcal{L}_{I_{X_R}}(s)\mathcal{L}_{I_{X_D}}(b)\rho_{X}(s,b)
\end{equation}

 \begin{equation} \label{eq26.2}
 \mathcal{L}_{I_{Y_R},I_{Y_D}}(s,b)=\mathcal{L}_{I_{Y_R}}(s)\mathcal{L}_{I_{Y_D}}(b)\rho_{Y}(s,b)
 \end{equation}
 
\begin{equation} \label{eq26.3}
\rho _{X}(s,b)=\exp\Bigg(p\lambda_{X}\int_{\mathcal{B}}\dfrac{sbP^2 l_{Rx} l_{Dx}}{\big(1+sP l_{Rx}\big)\big(1+bP l_{Dx}\big)}dx\Bigg)
\end{equation}
\begin{equation}\label{eq26.4}
\rho_{Y}\big(s,b\big)=\exp\Bigg(p\lambda_{Y}\int_{\mathcal{B}}\dfrac{sbP^2 l_{Ry} l_{Dy}}{\big(1+sP l_{Ry}\big)\big(1+bP l_{Dy}\big)}dy\Bigg).
\end{equation}

   \textit{Proof}:  See \ref{appE}.\hfill $ \blacksquare $\\
The cross term $ \rho_{X}(s,b) $  and $ \rho_{Y}(s,b) $ arise from the dependence between the interference at two locations (the relay and the destination). The integrals inside (\ref{eq26.3}) and (\ref{eq26.4}) can easily be calculated numerically with MATLAB software or Wolfram Mathematica.  Closed form can be obtained for $\alpha=2$ and $\alpha=4$.

 \section{Laplace Transform Expressions}
After we obtained the expressions of the outage probability, we derive, in this section, the Laplace transform expressions of the interference from the X road and from the Y road when the interference are originated from vehicles in finite road segments ($\mathcal{B}=[-Z,Z]$), and infinite road segments ($\mathcal{B}=\mathbb{R}$).
As mentioned in the previous section, joint Laplace transforms can be expressed as the product of two Laplace transforms and a cross term. 

The Laplace transform of the interference originating from the X road at the received node denoted  $N$, is expressed as

\begin{equation}\label{eq27} 
\mathcal{L}_{I_{X_N}}(s)=\exp\Bigg(-\emph{p}\lambda_{X}\int_\mathcal{B}\dfrac{1}{1+\big(A\Vert \textit{x}-N \Vert^\alpha\big)/sP}dx\Bigg),
\end{equation}
and
\begin{equation}\label{eq28} 
\Vert \textit{x}-N \Vert=\sqrt{\big(n\sin(\theta_{N})\big)^2+\big(x-n \cos(\theta_N) \big)^2 }.
\end{equation}

The Laplace transform of the interference originating from the Y road is given by

\begin{equation}\label{eq29} 
\mathcal{L}_{I_{Y_N}}(s)=\exp\Bigg(-\emph{p}\lambda_{Y}\int_\mathcal{B}\dfrac{1}{1+\big(A\Vert \textit{y}-N \Vert^\alpha\big)/sP}dy\Bigg),
\end{equation}
where
\begin{equation}\label{eq30} 
\Vert \textit{y}-N \Vert=\sqrt{\big(n\cos(\theta_{N})\big)^2+\big(y-n \sin(\theta_N) \big)^2 },
\end{equation}
where $n$ and $\theta_N $ are  the distance between the node $N$ and the intersection, and the angle between the node $N$ and the X road, respectively.

The expressions (\ref{eq27}) and (\ref{eq29}) can easily be calculated numerically with mathematical software as MATLAB or Mathematica.  Closed form can be obtained for $\alpha=2$ and $\alpha=4$ when $\mathcal{B}=\mathbb{R}$, and $\mathcal{B}=[-Z,Z]$. Due to lack of space, we present only the case when $\alpha=2$.

\textbf{Proposition 1}. \textit{The Laplace transform expressions of the interference at the node $N$  for an intersection scenario, when $\mathcal{B}=\mathbb{R}$, and when $\alpha=2$ are given by
}
\begin{equation}\label{eq31} 
\mathcal{L}_{I_{X_N}}(s)=\exp\Bigg(-\emph{p}\lambda_{X}\dfrac{sP}{A^{2}}\dfrac{\pi}{\sqrt{\big(n\sin(\theta_{N})\big)^2+sP/A^{2}}}\Bigg),
\end{equation}
\textit{and}
\begin{equation}\label{eq32} 
\mathcal{L}_{I_{Y_N}}(s)=\exp\Bigg(-\emph{p}\lambda_{Y}\dfrac{sP}{A^{2}}\dfrac{\pi}{\sqrt{\big(n\cos(\theta_{N})\big)^2+sP/A^{2}}}\Bigg).
\end{equation}

\textit{Proof}:  See \ref{appF}.\hfill $ \blacksquare $ \\
When a cooperative transmission is considered, $N\in\{R,D\}$ and $n\in\{r,d\}$.\\
Note that when $ \theta_N=0 $, it corresponds to the case where the received node is on the X  road. Then, if we substitute $ \theta_N=0 $ in (\ref{eq31}) and (\ref{eq32}), we obtain the special case in \cite{steinmetz2015stochastic} when the receiving node is on the X road, and when the direct transmission and infinite road segments are considered.

\textbf{Proposition 2}. \textit{The Laplace transform expressions of the interference at the node $N$  for an intersection scenario, when $\mathcal{B}=[-Z,Z]$ and when $\alpha=2$ are given by
}
\begin{equation}\label{eq33} 
\mathcal{L}_{I_{X_N}}(s)=\exp\big(-\emph{p}\lambda_{X} \Gamma_X(s) \big),
\end{equation}
and
\begin{equation}\label{eq34} 
\mathcal{L}_{I_{Y_N}}(s)=\exp\big(-\emph{p}\lambda_{Y}\Gamma_Y(s) \big),
\end{equation}
where
\begin{equation}\label{eq35} 
\Gamma_X(s)=\dfrac{\arctan\Bigg( \dfrac{Z+n_x}{\sqrt{n_y^{2}+sP/A^{2}}}\Bigg)+\arctan\Bigg( \dfrac{Z-n_x}{\sqrt{n_y^{2}+sP/A^{2}}}\Bigg)}{\dfrac{A^{2}}{sP}\sqrt{n_y^{2}+\dfrac{sP}{A^{2}}}},
\end{equation}
and
\begin{equation} \label{eq36} 
\Gamma_Y(s)=\dfrac{\arctan\Bigg( \dfrac{Z+n_y}{\sqrt{n_x^{2}+sPA^{2}}}\Bigg)+\arctan\Bigg( \dfrac{Z-n_y}{\sqrt{n_x^{2}+sPA^{2}}}\Bigg)}{\dfrac{A^2}{sP}\sqrt{n_x^{2}+\dfrac{sP}{A^{2}}}}.
    \end{equation}

\textit{Proof}:  See \ref{appG}.\hfill $ \blacksquare $ \\

It is worth noting that the expression of (\ref{eq31}) does not depend on $ n\cos(\theta_N) $, that is $ n_x$. Similarly, (\ref{eq32}) does not depend on $n\sin(\theta_N )$, that is $n_y$. However, one can notice  that (\ref{eq35}) and (\ref{eq36}) both depend on $n_x$ and on $n_y$. This can be explained by the fact that, as interferers tend to infinity, the term $(x-n_x)$ tends to $x$ and $(y-n_y)$ tends to $y$, that is $\lim\limits_{x \rightarrow \infty} (x-n_x)\longrightarrow x$, and $\lim\limits_{x \rightarrow \infty} (y-n_y)\longrightarrow y$. 
This is no longer the case when the interferers are on finite road segments. In this case, the result depends on $n_x$ and on $n_y$.

Now that the Laplace transform equations are at hand, we plug them into the outage probability equations, and we get the final and complete expressions of the outage probabilities.
\section{Simulations and Discussions}

\begin{figure*}[t!]
    \centering
    \begin{subfigure}[b]{0.5\textwidth}
        \centering
        \includegraphics[height=8cm,width=8cm]{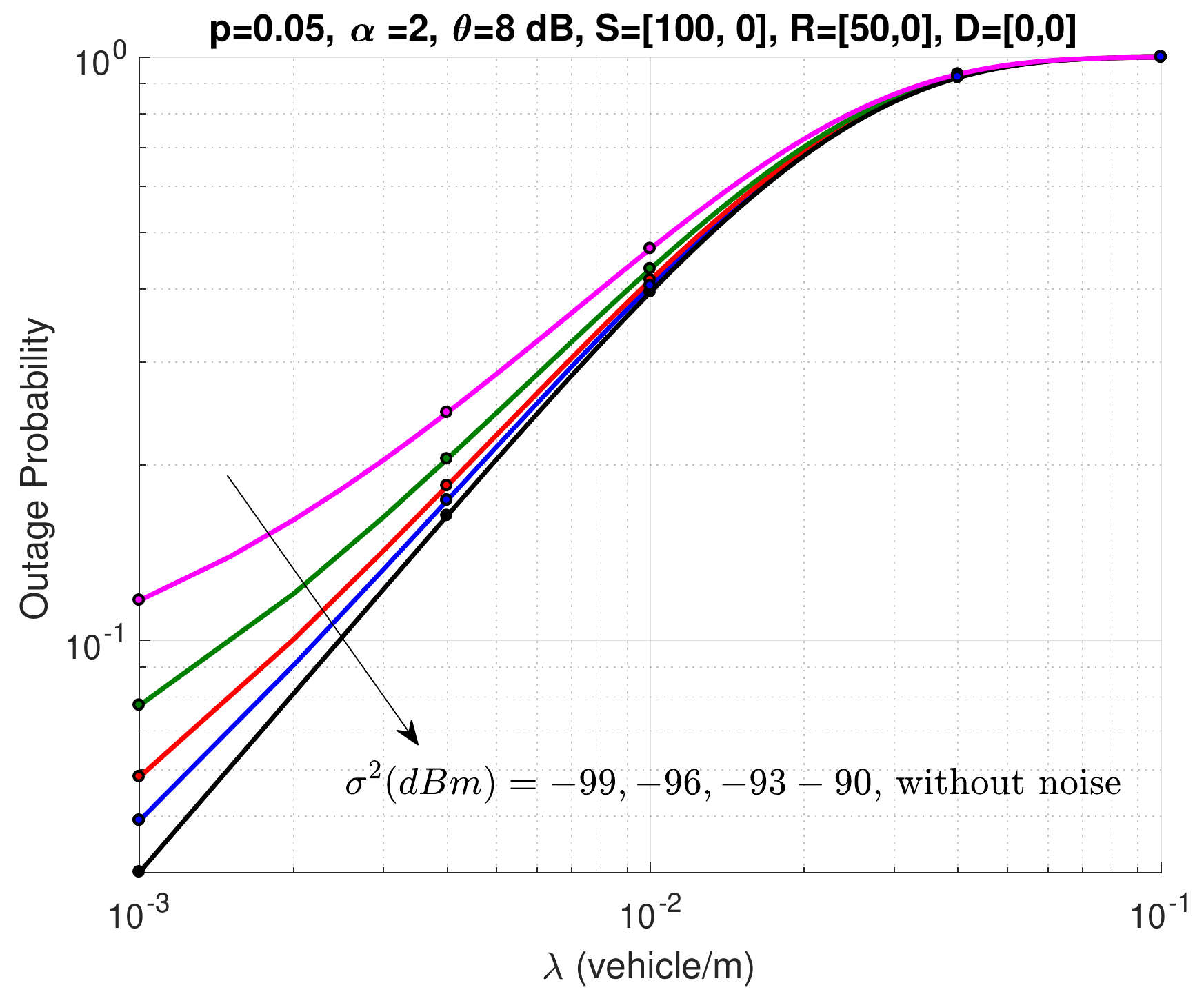}
        \caption{}
     \label{Fig2(a)}
    \end{subfigure}%
    ~ 
    \begin{subfigure}[b]{0.5\textwidth}
        \centering
        \includegraphics[height=8cm,width=8cm]{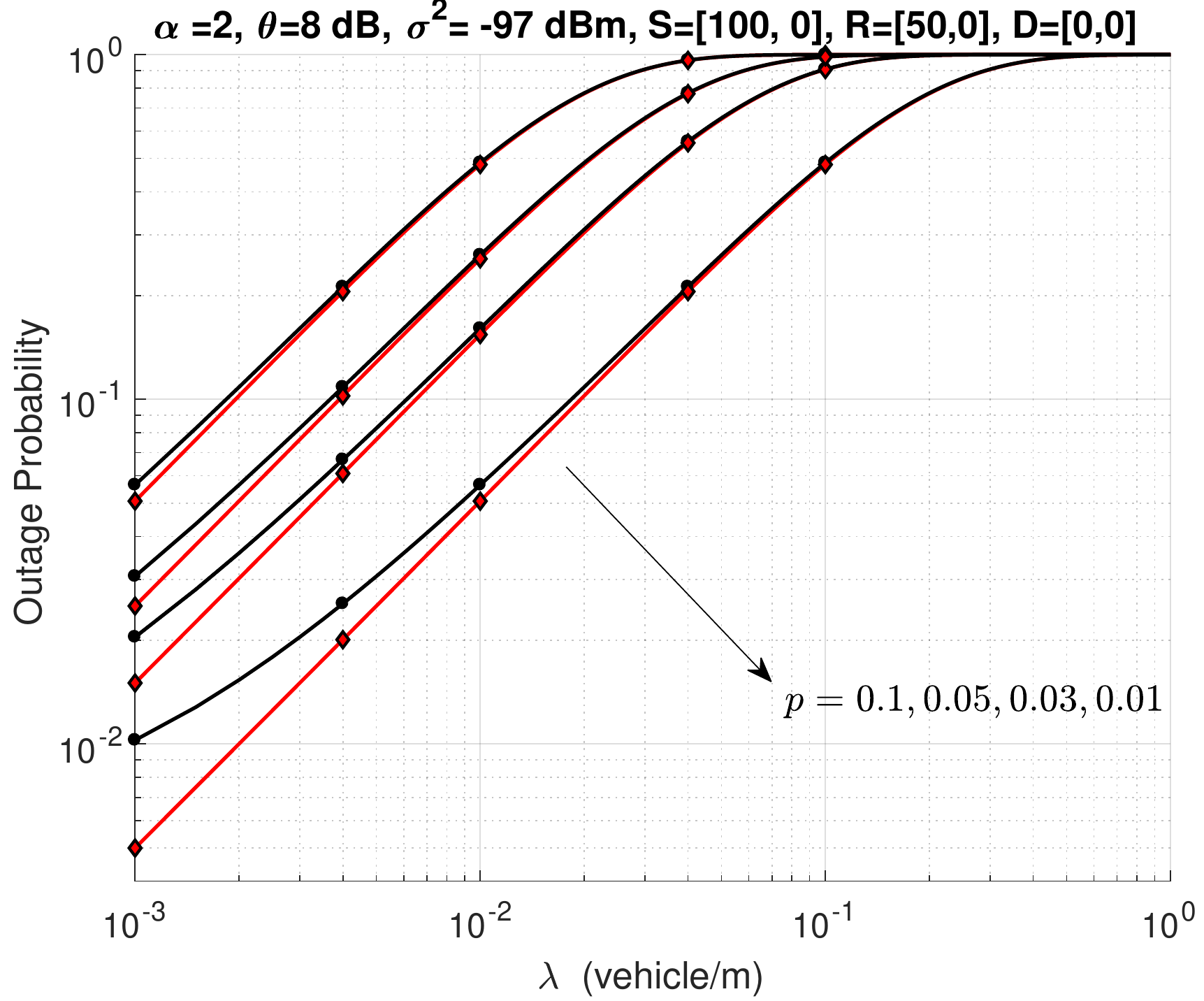}
        \caption{}
        \label{Fig2(b)}
    \end{subfigure}
    \caption{Outage probability when the vehicles intensity $\lambda$ considering SC and the HSV model. (a) represents the outage probability for several values of noise power level. (b) represents the outage probability for several values of $p$ with noise (circle), and without noise (diamond).}
    \label{Fig2}
\end{figure*}
In this section, we evaluate the performance of our model. In order to verify the accuracy of the theoretical results, Monte Carlo simulations are obtained by averaging over 10,000 realizations of the PPPs and fading parameters. Unless stated otherwise, all the figures are plot for intersection scenarios. We set our simulation area to [$-10^6$ m, $10^6$ m] for each road segment. We also set $A=650$, and the transmit power to $P =120$ mW. Without loss of generality, we set $\lambda_X=\lambda_Y=\lambda$. The vehicles intensity $\lambda$ can also be interpreted as the average distance between vehicles.\\
\subsection{Noise analysis}
Fig.\ref{Fig2(a)} plots the outage probability for several values of noise power level when $p=0.05$. For low interference level (low vehicle intensity), the noise becomes predominant, and thus degrades the performance. However, as the number of interfering vehicles increases, the noise becomes negligible. Fig.\ref{Fig2(b)} depicts the outage probability for several values of $p$ when the noise power level is set to $\sigma^2=-97$ dBm. We can notice that, as $p$ increases, the performance of the outage with noise and without noise converge to the same values. This is because, as $p$ increases, the power of interference increases accordingly, thus making the power of noise negligible compared the interference power, which corresponds to the interference limited scenario.\\
In the next figures, since we are mainly interested in the effect of the interference, we will consider only the interference limited scenario, that is, $\sigma^2=0$.
\begin{figure}[]
\centering
\includegraphics[scale=0.6]{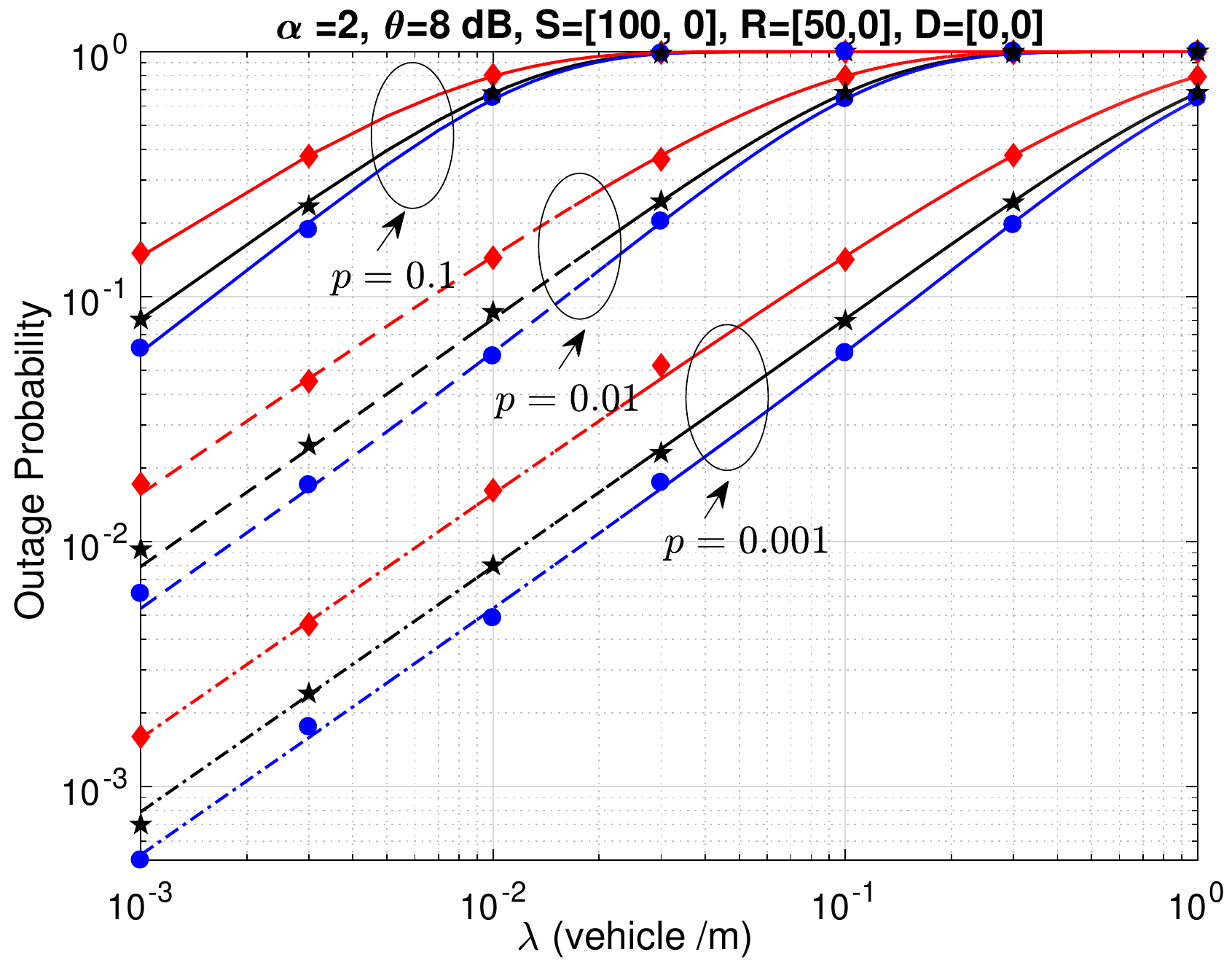}
\caption{Outage probability when varying $\lambda$ for different values of $p$ using the direct transmission scenario (diamond), and cooperative transmissions considering SC (star)  and MRC (circle), for the HSV model.}
\label{Fig3}
\end{figure}
\textit{}
\subsection{Mobility and decoding strategies analysis}

From Fig.\ref{Fig3}, we can make the following observations. 
First, we retrieve that MRC ouperforms SC. Second, cooperative transmissions outperform the direct transmission, that is, the outage probability for a cooperative transmission is lower than the one for a direct transmission. The explanation of the second observation is that, when the direct transmission fails, i.e., the link \textit{S-D} is in outage, it is unlikely that \textit{S-R} and \textit{R-D} are in outage too. Thus, the direct transmission is aided by the relaying paths \textit{S-R} and \textit{R-D}, and therefore the cooperative transmission always enhance the performance compared to the direct transmission. We notice that, for lower values of $p$, the outage probability decreases. This is because lower values of $p$ mean lower probability for the vehicles to access the medium, leading to less interferers, and thus reducing the SINR and the outage probability. We can conclude that vehicles should always cooperate in order to minimize the outage probability, and we confirm the superiority of MRC over SC. We study the effect of MRC in \ref{Fig6}.
\begin{figure*}[t!]
    \centering
    \begin{subfigure}[b]{0.5\textwidth}
        \centering
        \includegraphics[height=8cm,width=8cm]{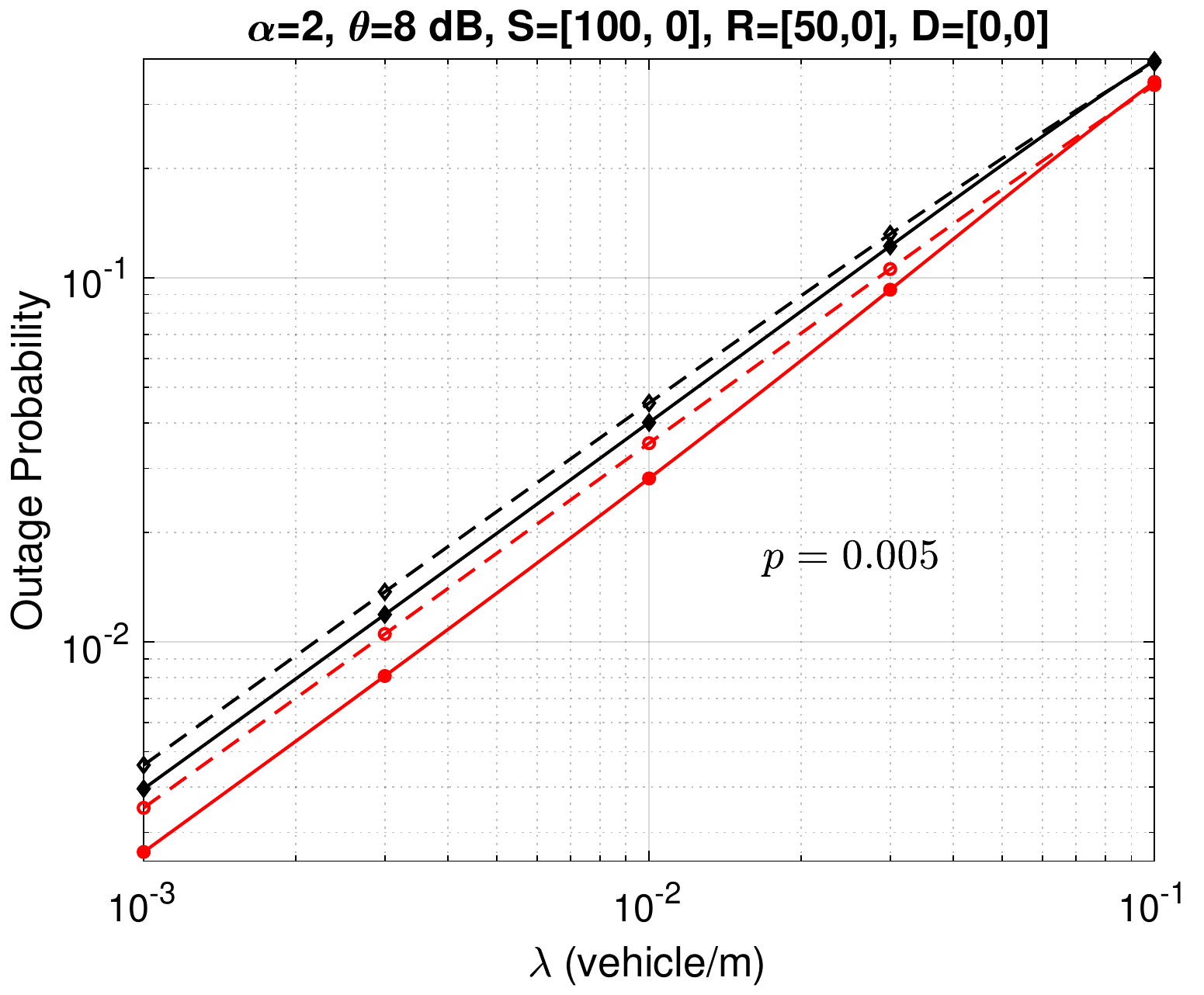}
        \caption{Low traffic scenario}
     \label{Fig4(a)}
    \end{subfigure}%
    ~ 
    \begin{subfigure}[b]{0.5\textwidth}
        \centering
        \includegraphics[height=8cm,width=8cm]{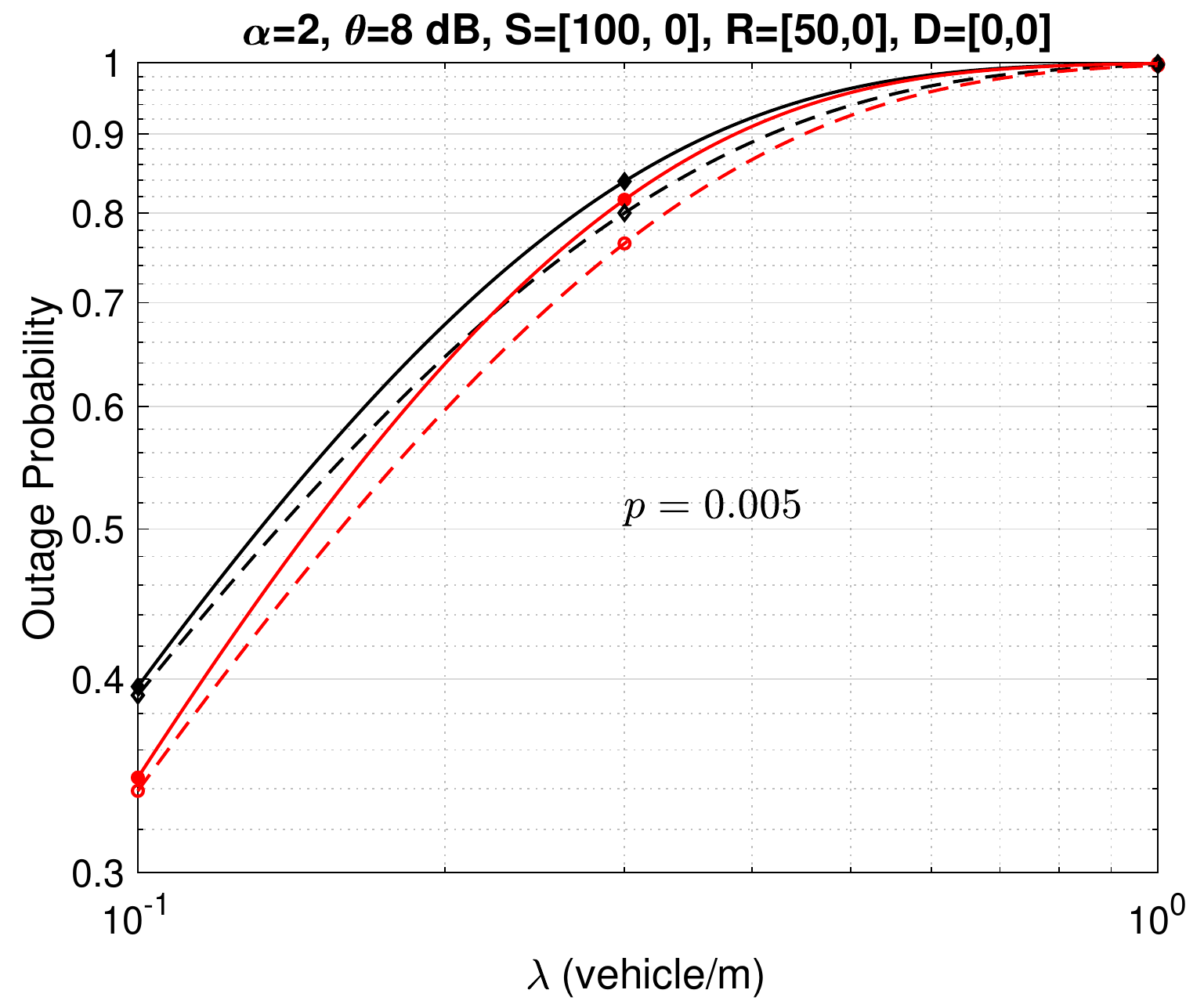}
        \caption{High traffic scenario}
     \label{Fig4(b)}    
    \end{subfigure}
    \caption{ Outage probability as a function of $\lambda$ considering SC (diamond), and MRC (circle), for the HSV model (simple line) and the LSV model (dashed line).}
  \label{Fig4}   
\end{figure*}

We notice from Fig.\ref{Fig4(a)} that, for lower values of $\lambda$ (low traffic conditions), the HSV model outperforms the LSV model. But we can notice from Fig.\ref{Fig4(b)}, as the intensity of vehicles increases (high traffic conditions), the LSV model exhibits better performance. This is explained by the fact that the interference dependence (LSV) in high traffic is beneficial due to highly dependent hops of the relaying path. So, if the \textit{S-R} link succeeded, it is likely to be the case of the \textit{R-D} link. Thus, the interference dependence (LSV) leads to higher performance than in the presence of independent interference (HSV) in high traffic conditions, that is, when the number of transmitting vehicles increases \cite{ganti2009spatial}. In other words, in low traffic conditions, increasing the vehicle speed increases the outage performance, whereas in high traffic conditions, decreasing the vehicle speed increases the performance.\\
\begin{figure}
\centering
\includegraphics[scale=0.6]{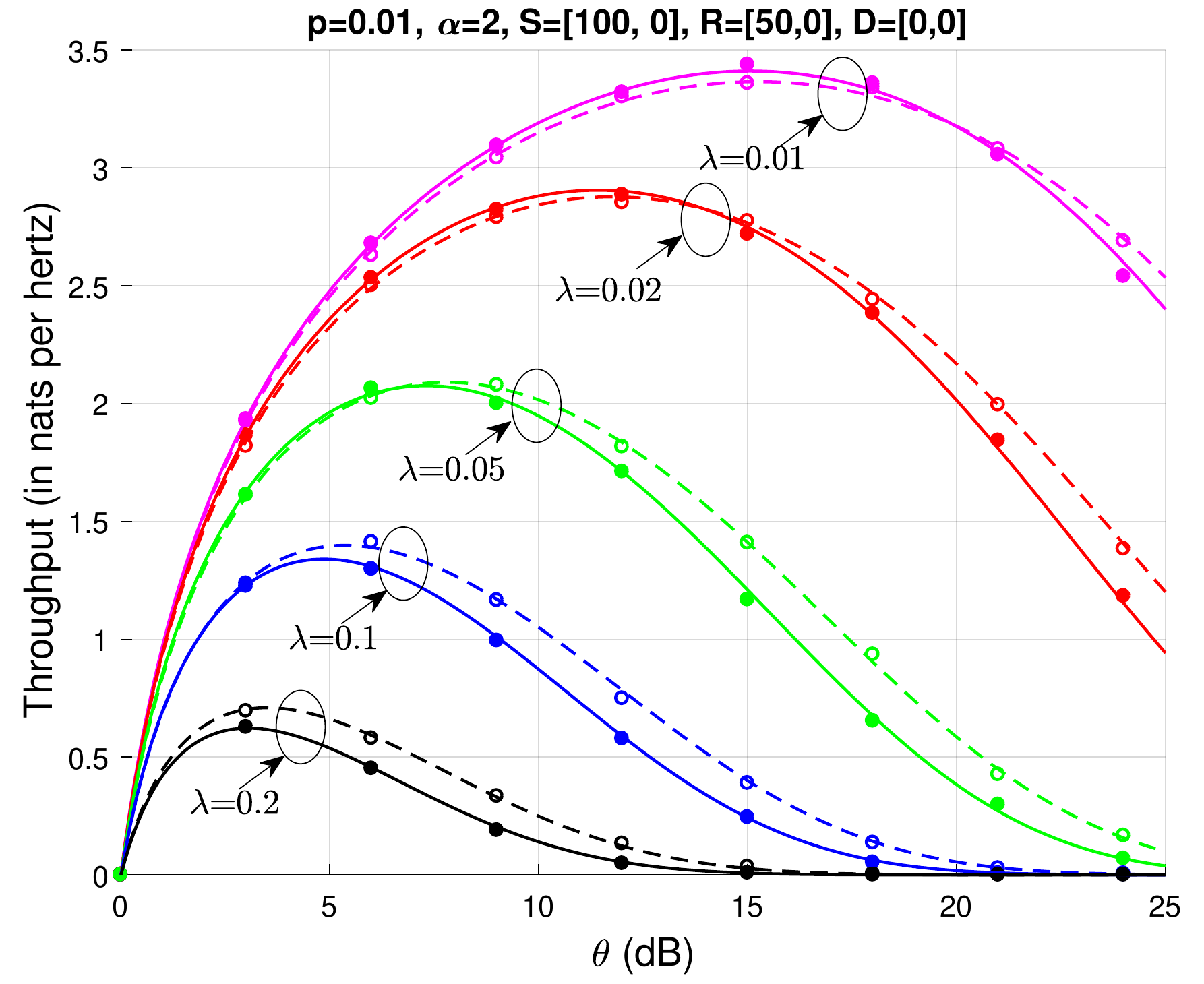} 
%\includegraphics[height=7.5cm,width=8cm]{Fig2.pdf} 
%\begin{flushleft}
\caption{Throughput as a function of $\Theta$ for different values of $\lambda$ considering MRC, for the HSV model (simple line) and the LSV model (dashed line). Simulation results are represented with dots.}
 \label{Fig5}
\end{figure}
In Fig.\ref{Fig5}, we plot the throughput as a function of $\Theta$, where the throughput $\mathcal{T}$ is defined as follows
\begin{equation}
\mathcal{T}=\mathbb{P}(O^C)\log_2(1+\Theta),\nonumber   
\end{equation}       
where $\log_2(1+\Theta)$  is the Shannon bound (in nats pers hertz) and $\mathbb{P}(O^C)$ is the success probability \cite{haenggi2009outage}. 
We can notice that in a high traffic scenario $(\lambda=0.1,\ \lambda=0.2)$, the LSV model allows the highest throughput than the HSV model. This confirms what we concluded in Fig.\ref{Fig4}. We also notice that, in a low traffic scenario $(\lambda=0.01,\ \lambda=0.02)$, the HSV model allows a slightly higher throughput than the LSV model. However, even for lower values of $\lambda$ (low traffic), as $\Theta$ increases, the LSV model achieves a higher throughput that the HSV model. This is because, for larger values of  $\Theta$, in order to have an outage, a large number of vehicles have to transmit at the same time, hence, increases the traffic density.  

Note that $\mathcal{T}$ is a function of $\log_2(1+\Theta)$ and $\mathbb{P}(O^C)$. When $\Theta$ increases, $\log_2(1+\Theta)$ increases whereas the success probability $\mathbb{P}(O^C)$ decreases.  In one hand, we are tempted to increase $\Theta$ to increase the rate; but, on the other hand, increasing $\Theta$ increases the outage probability. An optimal value of $\Theta$ must be carefully set in order to maximize the throughput under given traffic conditions and vehicles mobility.\\

In Fig.\ref{Fig6}, we plot the outage probability as a function of the relay position for a setting where $S=(0,0)$ and $D=(200, 0)$. We can see from Fig.\ref{Fig6(a)} that the best relay position for the LSV model is in the middle of \textit{S} and \textit{D}, whereas the best relay position for the HSV model is slightly shifted from the middle toward \textit{D}. We also can see from Fig.\ref{Fig6(a)} that, in high traffic scenarios, the LSV model achieves a better performance (as stated in Fig.\ref{Fig4}). However, when the relay is close to the destination, the HSV model has a better performance than the LSV model. This is because, in a high traffic scenario (harsh environment), the direct link \textit{S-D} is more likely to be in outage, therefore when the relay is close to the destination, that is, $\Vert S-R \Vert \approx \Vert S-D \Vert$, the \textit{S-R} link is more likely to be in outage due to highly dependent interference.
\begin{figure*}[t!]
    \centering
    \begin{subfigure}[b]{0.5\textwidth}
        \centering
        \includegraphics[height=8cm,width=8cm]{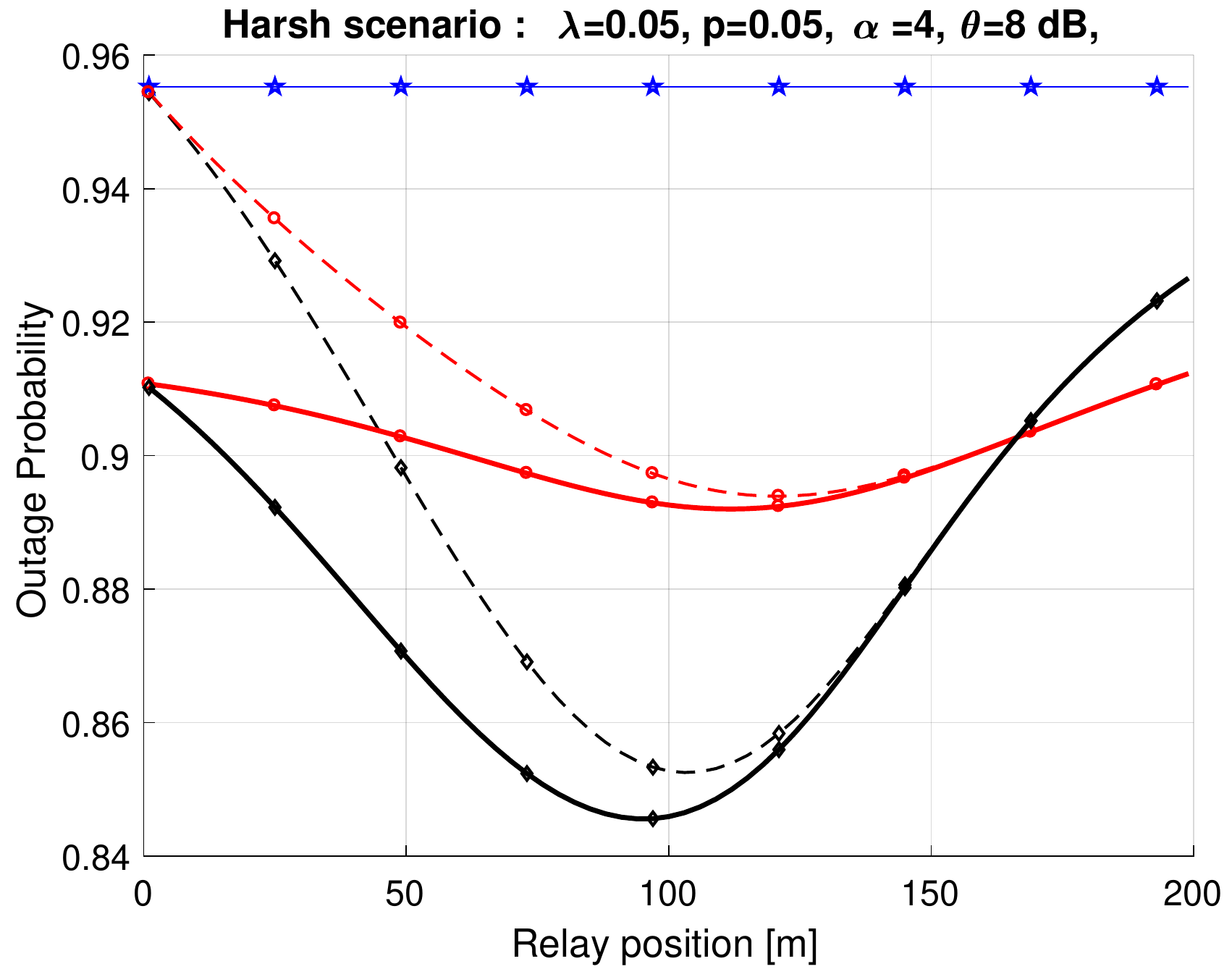}
        \caption{High traffic scenario: $\lambda=0.25$}
       \label{Fig6(a)}  
    \end{subfigure}%
    ~ 
    \begin{subfigure}[b]{0.5\textwidth}
        \centering
        \includegraphics[height=8cm,width=8cm]{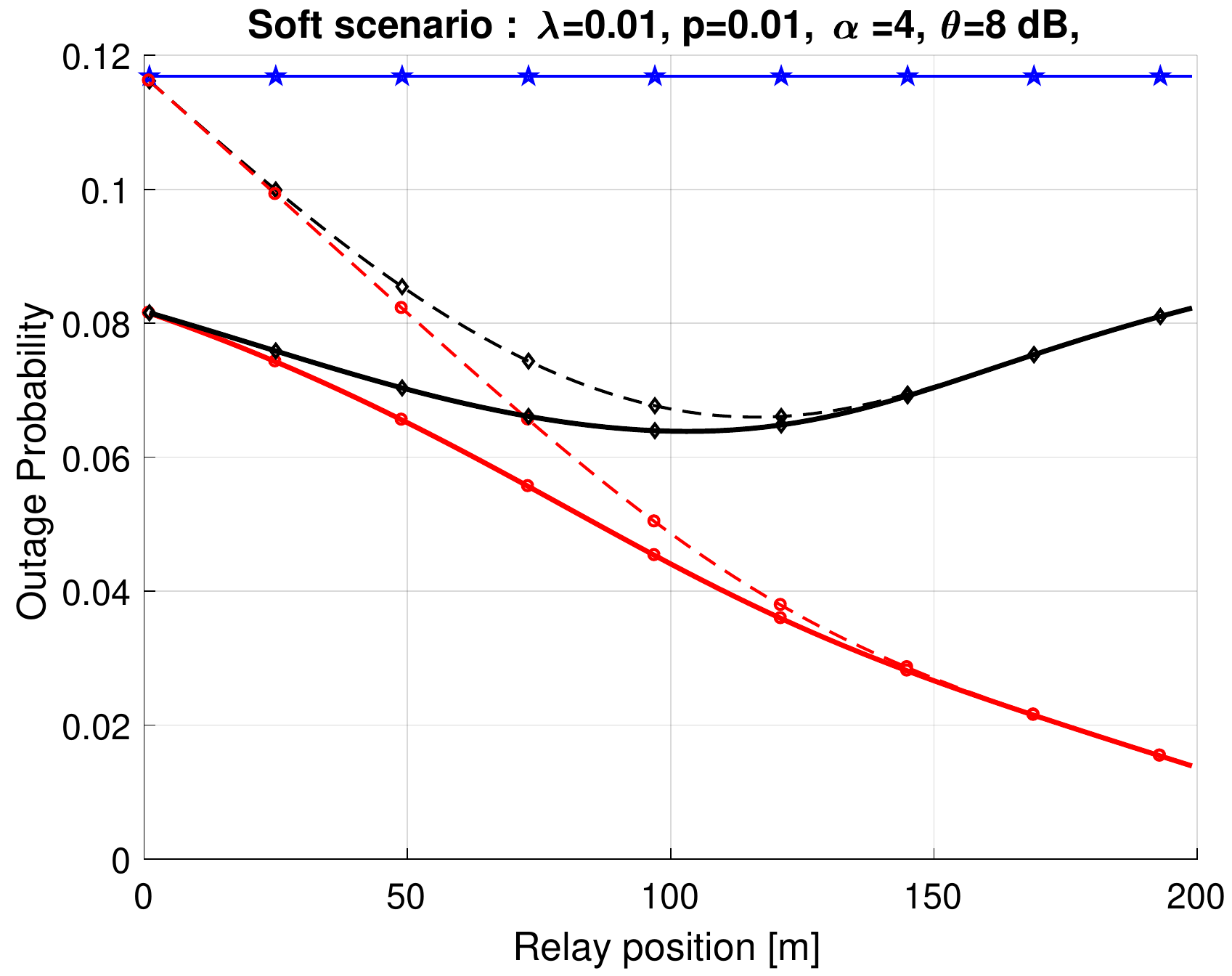}
        \caption{Low traffic scenario: $\lambda=0.01$}
      \label{Fig6(b)}   
    \end{subfigure}
    \caption{ Outage probability as a function of the relay position, considering SC (dashed line), MRC (line) and the direct transmission (star), for the HSV model (circle) and the LSV model (diamond).}
    \label{Fig6} 
\end{figure*}
Furthermore, as stated for Fig.\ref{Fig4}, in low traffic scenario, the HSV model has a better performance than the LSV model. We can notice from Fig.\ref{Fig6(b)} that the best relay position for the HSV model is close to the destination whereas the best relay position for the LSV model is when the relay is equidistant from \textit{S} and \textit{D}. The explanation is as follow: in low traffic scenarios, the direct link has a high success probability in the presence of low interference level. However, in the HSV model, even if the direct link fails, it is less likely that the relay path fails too, since there is no dependence between interference. Hence, when the relay moves toward the destination, it increases the diversity gain and enhances the performance. This makes the best relay position in the HSV model close to the destination. However, in the LSV model, when the direct link fails, it is more likely that the relay path fails due to (low but still present) interference dependence. Hence, the best relay position is when the relay is equidistant from \textit{S} and \textit{D}.\\
\begin{figure*}[t!]
    \centering
    \begin{subfigure}[b]{0.5\textwidth}
        \centering
        \includegraphics[height=8cm,width=8cm]{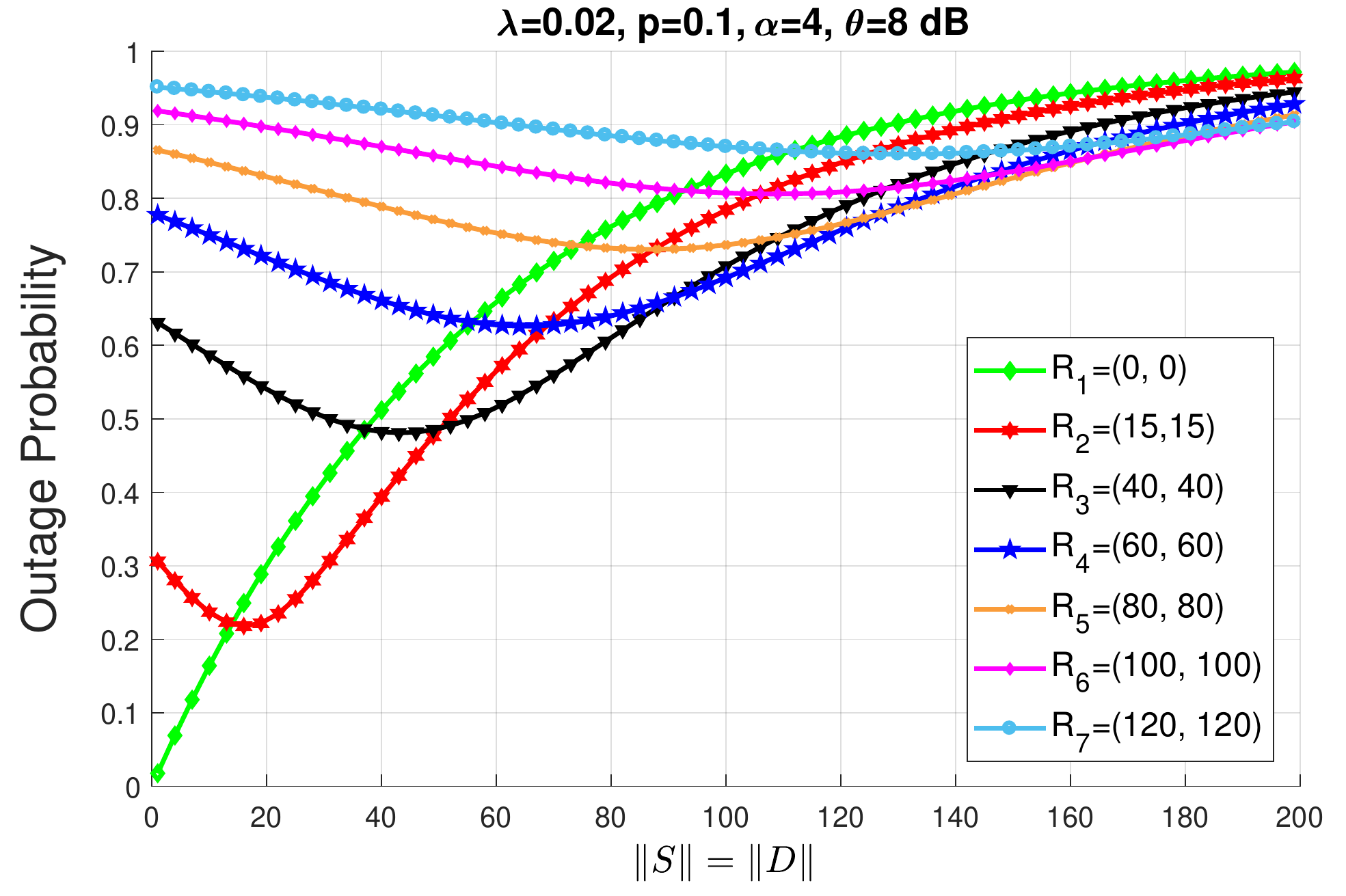}
        \caption{2D plot}
     \label{Fig7(a)}    
    \end{subfigure}%
    ~ 
    \begin{subfigure}[b]{0.5\textwidth}
        \centering
        \includegraphics[height=8cm,width=8cm]{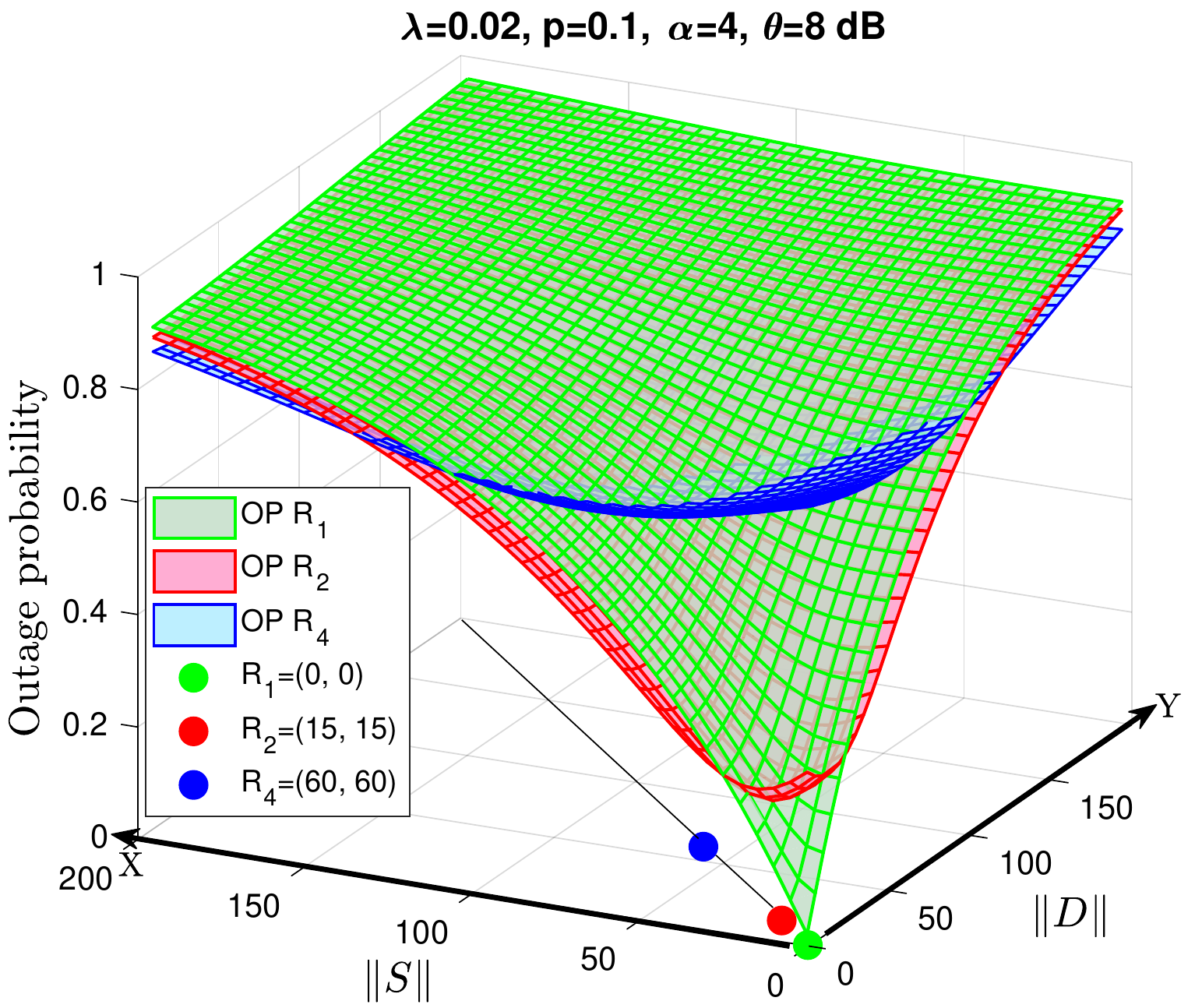}
        \caption{3D plot}
      \label{Fig7(b)}   
    \end{subfigure}
    \caption{ Outage probability when varying the distance of the destination and the source from the intersection, denoted respectively $\Vert S \Vert$ and $\Vert D \Vert$ for several locations of the relay at the first bisector, in the LSV model. (a) represents the 2 dimensions plot, and (b) represents the 3 dimensions plot. }
     \label{Fig7}
\end{figure*}
Finally, we can notice, regardless of traffic conditions or vehicle speeds that, as the relay moves closer to the destination, MRC and SC offer the same performance. This is because, when the relay is close to the destination, the power received at the destination from the source is much smaller than the power received from the relays ($l_{RD} \gg l_{SD}$). Thus, adding the power of \textit{S-D} link does not add much power to the \textit{R-D} link, which makes MRC and SC at the same level of performance. When the relay is closer to the source, the level of power received at the destination from source and from the relay is almost the same ($l_{RD} \approx l_{SD}$), which increases the diversity gain, leading to greater performance of MRC over SC.\\
The relay location plays an important role in the performance. This can be used in the relay-selection based algorithms, where vehicles have to take into account both the relays location and speeds (HSV or LSV). Regarding the decoding strategies, there is also a tradeoff between performance and complexity to consider, because MRC is difficult to implement, and it is only beneficial when the relay is close to the source.
\begin{figure*}[t!]
    \centering
    \begin{subfigure}[b]{0.5\textwidth}
        \centering
        \includegraphics[height=8cm,width=8cm]{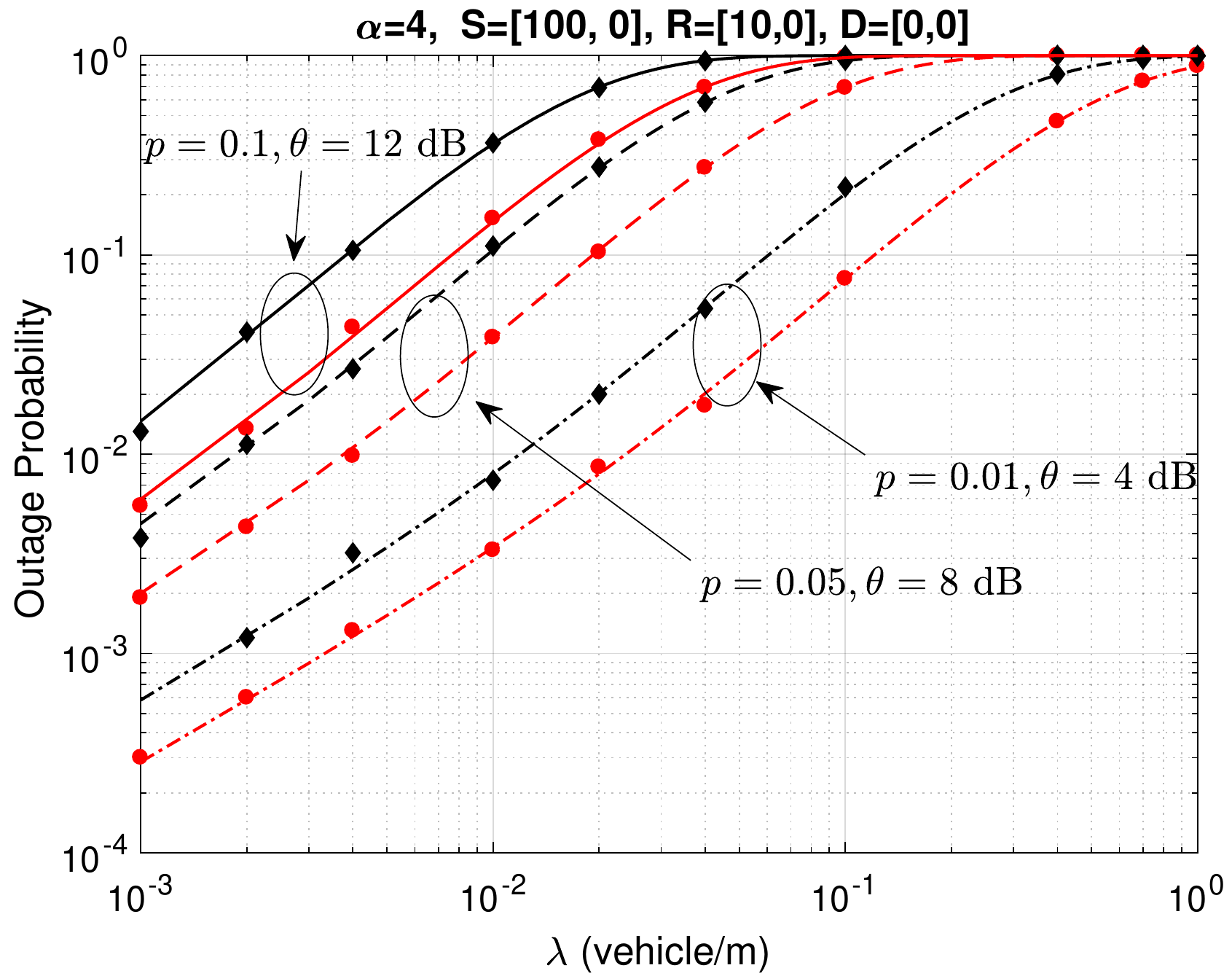}
        \caption{}
       \label{Fig8(a)}  
    \end{subfigure}%
    ~ 
    \begin{subfigure}[b]{0.5\textwidth}
        \centering
        \includegraphics[height=8cm,width=8cm]{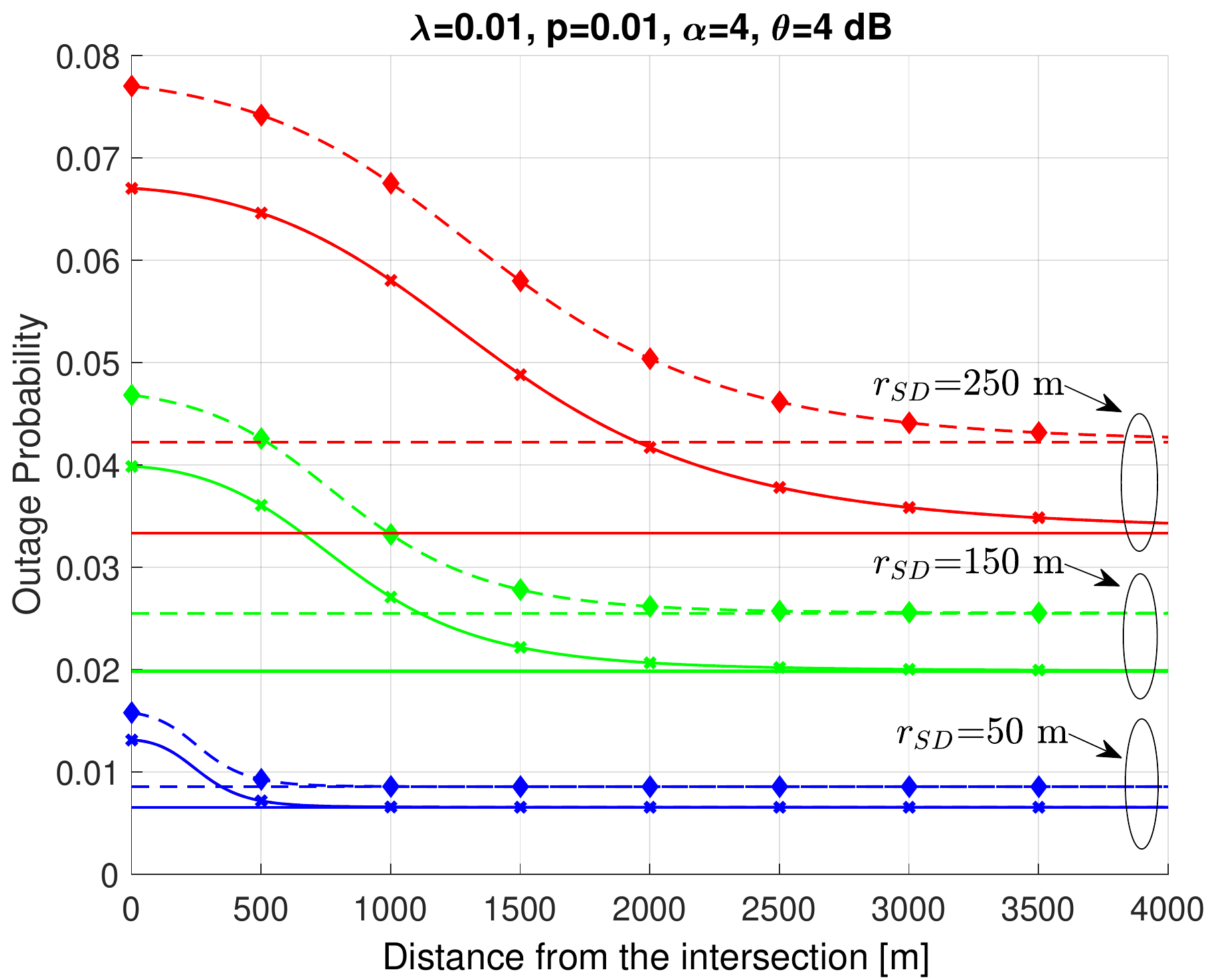}
        \caption{}
       \label{Fig8(b)}  
    \end{subfigure}
    \caption{(a) Outage probability as function of $\lambda$ for different values of $p$ and $\Theta$ in a highway scenario (circle) and intersection scenario (diamond) considering MRC for the HSV model. Analytical results are plot with lines, and simulation results with marks.(b) Outage probability when varying the distance of the triplet $\{S,R,D\}$ from the intersection for different values of $r_{SD}$. The highway scenario (line without marks) and the intersection scenario (line with marks) are considered, for the HSV model (simple line) and the LSV model (dashed line).}
   \label{Fig8}  
\end{figure*}

\subsection{Intersection and highway scenario analysis}
In Fig.\ref{Fig7}, we present analytical results when the relay is located on the first bisector. We plot the outage probability as a function of the distance of the source from the intersection, and the distance of the destination from the intersection, denoted respectively $\Vert S \Vert$ and $\Vert D \Vert$, in 2D in Fig.\ref{Fig7(a)}, and in 3D in Fig.\ref{Fig7(b)}. Without loss of generality, we set the source on the X road, and the destination on the Y road.\\
Since the relay is outside the roads, we can consider that it belongs to the roadside infrastructure. A roadside infrastructure with the coordinate $(0,0)$ can be placed in the center of a roundabout, or mounted on a traffic light pole. We can notice that the outage probability reaches it minimum when the relay is the middle of \textit{S} and \textit{D}. We also notice that there is no need to use a relay that would be farther than the coordinate (60, 60) in terms of outage performance. Although there is a little bit of (but still negligible) gain when using $R_3$, we can state that using only 3 relays in 200 meters or above, offer the same performance than using 7 relays or more.
In Fig.\ref{Fig7(b)}, we only plot the outage probability when using the relay $R_1$, $R_2$ and $R_4$. We can notice that the relay $R_1$ covers the first [0,17 m] x [0,17 m], that is, 289 m$^2$ (300 m$^2$). The relay $R_2$ cover approximately 4600 m$^2$ (4611 m$^2$), and then the relay $R_4$ cover the rest, that is, more than 35000 m$^2$ (35389 m$^2$).\\
\begin{figure}
\centering
\includegraphics[scale=0.56]{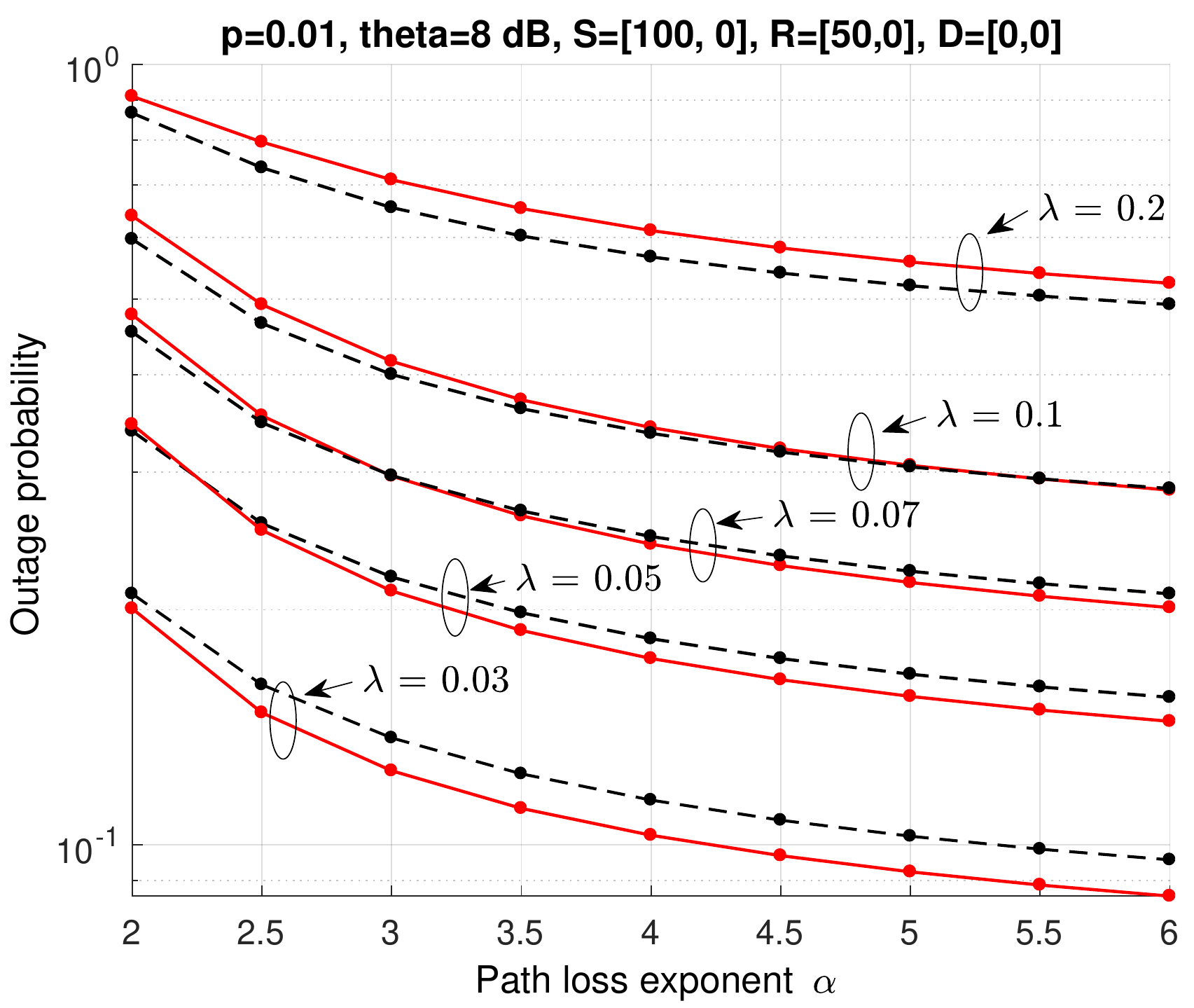} 
%\includegraphics[height=7.5cm,width=8cm]{Fig2.pdf} 
%\begin{flushleft}
\caption{Outage probability as a function of the path loss exponent $\alpha$ for $\lambda \in \{0.03, 0.05, 0.07, 0.1, 0.2\}$, for the HSV model (line) and the LSV model (dashed line) considering MRC.}
 \label{Fig9}
\end{figure}
Fig.\ref{Fig8(a)} compares the outage probability of a cooperative highway scenario with a single lane \cite{farooq2016stochastic} and a cooperative intersection scenario with two orthogonal lanes. We can see from Fig.\ref{Fig8} that the outage probability increases as $\lambda$, $p$ and $\Theta$ increase (see Fig.\ref{Fig3}). One can notice that the highway scenario outperforms the intersection scenario in terms of outage probability. We infer that the intersection scenario has an additional lane that contributes to the aggregate of interference, therefore increasing the outage probability.\\
In Fig.\ref{Fig8(b)}, we present the results for several values of $r_{SD}$, the relay is equidistant from the source and the destination. Without loss of generality, we set the triplet $\{S,R,D\}$ on the same road. We plot the outage probability as a function of the distance from the intersection for $r_{SD} \in \{$50 m, 150 m, 250 m$\}$.
As for the results  in Fig.\ref{Fig8(a)}, the highway scenario offers a better performance in terms of outage probability than the intersection scenario. But, as we increase the distance between the triplet and the intersection, the highway scenario and the intersection scenario converge toward the same value. This can be explained by the fact that, as vehicles move away from the intersection, the power of the interference originated from the other road becomes negligible, thus leading to the same performance as in a highway scenario. 

This further confirms the statement that the intersection scenario has a higher outage probability compared to the highway scenario, thus making intersections more critical areas because they are more prone to outage.

\subsection{Environment analysis}
From Fig.\ref{Fig9}, we see that, as $\alpha$ increases, the outage probability decreases, this is intuitive, since the path loss function decreases faster for larger value of $\alpha$, thus leading to the rapid decrease of the interference power. We also see that, in a high traffic load ($\lambda=0.2$), the LSV model exhibits better performance. Inversely, in low traffic load, the HSV model exhibits a better performance. This confirms the results in Fig.\ref{Fig2}, Fig.\ref{Fig3} and Fig.\ref{Fig6}.
We note that, for a low traffic scenario ($\lambda=0.03$, $\lambda=0.05$), the gap in terms of outage probability between the LSV model and the HSV model, for $\alpha=2$, is very small, but,  as $\alpha$ increases, the gap becomes larger. 
Also, for $\lambda=0.07$, the HSV model has a slightly higher value of outage probability than the LSV model when $\alpha=2$, but, as the value of $\alpha$ increases, the LSV model has a slightly high value of outage probability over the HSV model. 

We conclude that larger values of $\alpha$ lead to higher interference dependence, because the interference is dominated by vehicles that are close to the receiving nodes. On the opposite, lower values of $\alpha$ lead to lower interference dependence, because the interference is a summation of several far vehicle signals which, to some extent, decreases the dependence of the interference \cite{haenggi2012diversity}. Note that the path loss exponent $\alpha$ depends on the environment. For instance, $\alpha=2$ corresponds to the free space,  $\alpha \in \{2.7, 3.5\}$ to an urban area, and $\alpha \in \{3, 5\}$ to a shadowed urban area \cite{rappaport2002wireless}. We can state that the environment plays an important role in the interference dependence. Indeed, an intersection in a suburban area has a lower dependence of interference, whereas an intersection in an urban has a higher interference dependence.

\section{Conclusion}

In this paper, we studied the performance of direct transmissions and cooperative transmissions for vehicular networks at road intersections in the presence of interference. We presented analytical results for two mobility models. Closed-from expressions were obtained for specific channel conditions. We also considered two decoding strategies: SC and MRC. We derivate Laplace transform expressions when the receiving node can be anywhere on the plan, given finite and infinite road segments.
We showed that cooperative transmissions always enhance the performance compared to direct transmissions, and the use of MRC is only useful when the relay is closer to the source. We also showed that mobility increases the outage probability performance in good traffic conditions, whereas static or low mobility increases the outage probability performance in harsh traffic conditions. We also showed that the best relay position for the HSV model in low traffic conditions is when relay is closer to the destination. The best relay position for the HSV model in high traffic conditions is when relay is slightly shifted from the middle toward the destination. However, the best relay position in the LSV model is when the relay is equidistant from the source and the destination regardless of the traffic conditions. We showed that the outage probability does not improve after using three infrastructure relays. We also showed that lower values of the path loss exponent leads to higher interference dependence. Finally, we showed that cooperative transmissions at intersections have higher outage probability than cooperative transmissions on highways. 
As future works, we plan to analyze other  medium access protocols, such as CSMA. We also plan to consider other channel models and other roads geometries.

\appendix
\section{}\label{appA}
\begin{center}
PROOF OF LEMMA 1
\end{center}
We calculate the probability $\mathbb{P}(\textit{O}^{C}_{SR} \cap \textit{O}^{C}_{SD})$ in equation (\ref{eq14})
\begin{multline} \label{eq37} 
\mathbb{P}(\textit{O}^{C}_{SR} \cap \textit{O}^{C}_{SD}) =\mathbb{E}_{I_{X},I_Y}\Bigg[\mathbb{P}\Bigg\lbrace\Big(\dfrac{1}{2}\log_{2}\Bigg[1+\dfrac{P\vert h_{SR}\vert^{2}l_{SR}}{\sigma^{2}+I_{X_{R}}+I_{Y_{R}}}\Bigg]\ge R\\,\,\dfrac{1}{2}\log_{2}\Bigg[1+\dfrac{P\vert h_{SD}\vert^{2}l_{SD}}{\sigma^{2}+I_{X_{D}}+I_{Y_{D}}}\Bigg] \ge R\Bigg\rbrace\Bigg]\nonumber
\end{multline}
\begin{equation} \label{eq38} 
=\mathbb{E}_{I_{X},I_Y}\Bigg[\mathbb{P}\Bigg\lbrace\Big(\vert h_{SR}\vert^{2} \ge \dfrac{\Theta}{{Pl_{SR}}} (\sigma^{2}+I_{X_{R}}+I_{Y_{R}}),\,\vert h_{SD}\vert^{2}  \ge \dfrac{\Theta}{Pl_{SD}} (\sigma^{2}+I_{X_{R}}+I_{Y_{R}}) \Bigg\rbrace\Bigg].\nonumber
\end{equation}
Since $|h_{SR}|^2$ and $|h_{SD}|^2$ both follow an exponential distribution with unit mean, we get
\begin{multline} 
\mathbb{P}(\textit{O}^{C}_{SR} \cap \textit{O}^{C}_{SD}) =\mathbb{E}_{I_{X},I_Y}\Bigg[\exp(-K_{SR}I_{X_R})\exp(-K_{SR}I_{Y_R})\exp(-K_{SR}\sigma^{2})\\\exp(-K_{SD}I_{X_D})\exp(-K_{SD}I_{Y_D})
\exp(-K_{SD}\sigma^{2})\Bigg]\nonumber
\end{multline}
\begin{multline} 
=\mathbb{E}_{I_{X},I_Y}[N_{SR}\\ \exp(-K_{SR}I_{X_R})\exp(-K_{SR}I_{Y_R})N_{SD}\exp(-K_{SD}I_{X_D})\exp(-K_{SD}I_{Y_D})].\nonumber
\end{multline}
%where $Nr_{ab}=-exp(-K(r_{ab}\sigma^2)$. 
Given that the noise is independent of the interference, and using the independence of the PPP on the road $X$ and $Y$, we finally get (\ref{eq17}).

\section{}\label{appB}
\begin{center}
PROOF OF LEMMA 2
\end{center}To calculate the probability in $\mathbb{P}(O_{SR}^C\cap \textit{O}_{RD} )$, we proceed as follows
\begin{multline} 
\mathbb{P}(O_{SR}^C\cap \textit{O}_{RD} )=\\
\mathbb{E}_{I_{X},I_Y}\Bigg[\mathbb{P}\Bigg\lbrace\Big\vert h_{SR}\vert^{2} \ge \dfrac{\Theta}{Pl_{SR}}(\sigma^{2}+I_{X_{R}}+I_{Y_{R}}),\vert h_{RD}\vert^{2} < \dfrac{\Theta}{Pl_{RD}} (\sigma^{2}+I_{X_{D}}+I_{Y_{D}}) \Bigg\rbrace\Bigg].\nonumber
\end{multline}
\begin{multline} 
=\mathbb{E}_{I_{X},I_Y}\Bigg[\exp(-K_{SR}I_{X_R})\exp(-K_{SR}I_{Y_R})
\exp(-K_{SR}\sigma^{2})\\\ \Bigg(1-\exp(-K_{RD}I_{X_D})\exp(-K_{RD}I_{Y_D})\exp(-K_{RD}\sigma^{2})\Bigg)\Bigg]\nonumber
\end{multline}
\begin{multline} 
=N_{SR}\: \mathbb{E}_{I_{X},I_Y}\big[\exp(-K_{SR}I_{X_R}) 
\exp(-K_{SR}I_{Y_R})\big]\\-N_{SR}N_{RD}\: \mathbb{E}_{I_{X},I_Y}\big[\exp(-K_{SR}I_{X_R})\exp(-K_{SR}I_{Y_R})\exp(-K_{RD}I_{X_D})\exp(-K_{RD}I_{Y_D})\big].\nonumber
\end{multline}
\section{}\label{appC}
\begin{center}
PROOF OF LEMMA 3
\end{center}
To calculate the probability $\mathbb{P}(O_{SR}^C\cap \textit{O}_{SRD} )$, we follow the same steps as in Appendix A, then we obtain:
\begin{multline} \label{eq44} 
\mathbb{P}(O_{SR}^C\cap \textit{O}_{SRD} )=
\mathbb{E}_{I_{X},I_Y}\Bigg[N_{SR}\exp(-K_{SR}I_{X_R})\exp(-K_{SR}I_{Y_R})\\\ \Bigg[1-\mathbb{P}\Bigg\lbrace\vert h_{RD}\vert^{2}l_{RD}+\vert h_{SD}\vert^{2}l_{SD}\ge\dfrac{\Theta}{P}(\sigma^{2}+I_{X_{D}}+I_{Y_{D}})\Bigg\rbrace\Bigg]\Bigg].
\end{multline}
We write the probability inside the expectation in (\ref{eq44}) as
\begin{equation} \label{eq45} 
\mathbb{P}(\delta  \ge \beta[\sigma^2+I_{X_{D}}+I_{Y_{D}}]),\nonumber
\end{equation}
where $\delta=\vert h_{RD}\vert^{2}l_{RD}+\vert h_{SD}\vert^{2}l_{SD}$ and $\beta=\Theta/P$. \\
The complementary cumulative distribution function of the random variable $\delta$, denoted $\bar{F}_\delta(.)$, is given by
\begin{center} \label{eq46} 
$\bar{F}_\delta(u)=\dfrac{l_{RD} e^{-u/l_{RD}}-l_{SD} e^{-u/l_{SD}}}{l_{RD}-l_{SD}}$.
\end{center}
Then 
\begin{multline} \label{eq47} 
\mathbb{P}\big[\delta \ge \beta (\sigma^2+I_{X_D}+I_{Y_D})\big]=\\
\dfrac{l_{RD} \exp\Big[- \dfrac{\beta}{l_{RD}}(\sigma^2+I_{X_D}+I_{Y_D})\Big]- l_{SD}\exp\Big(- \dfrac{\beta}{l_{SD}}[\sigma^2+I_{X_D}+I_{Y_D})\Big]}{l_{RD}-l_{SD}},
\end{multline}
plugging (\ref{eq47}) into (\ref{eq44}), and with some algebraic manipulations, we get (\ref{eq19}).
%\begin{multline}\label{22}
%\mathbb{P}(\textit{O}^{C}_{SR} \cap \textit{O}^{C}_{SRD})=\dfrac{N_{SR}N_{RD}l_{RD}}{l_{RD}-l_{SD}} 
%\mathbb{E}_{I_X}[exp(-K_{SR}I_{X_R}-K_{RD}I_{X_D})] 
%\mathbb{E}_{I_Y}[exp(-K_{SR}I_{Y_R}-K_{RD}I_{Y_D})]\\ \nonumber
%- \dfrac{N_{SR}N_{SD}l_{SD}}{l_{RD}-l_{SD}} 
%\mathbb{E}_{I_X}[exp(-K_{SR}I_{X_R}-K_{SD})I_{X_D})]\mathbb{E}_{I_Y}[exp(-K_{SR}I_{Y_R}-K_{SD}I_{Y_D})].
%\end{multline}
\section{}\label{appD}
\begin{center}
PROOF OF THEOREM 1
\end{center}
When the interference at the relay and the destination are generated from two independent sets, the following expectation can be written as
\begin{equation} \label{eq48} 
\mathbb{E}_{I_X}\big[ e^{-(s I_{X_R}+b I_{X_D})}\big]=\mathbb{E}_{I_X}\big[ e^{-s I_{X_R}}\big]\mathbb{E}_{I_X}\big[ e^{-b I_{X_D}}\big].
\end{equation}
Given that  $\mathbb{E}[e^{sI}]=\mathcal{L}_I(s)$,  we then develop the expression of the first expectation in (\ref{eq48}) as
\begin{equation} \label{eq49} 
\mathcal{L}_{{I_{X_R}}}(s)=\mathbb{E}\big[{\\\exp(-sI_{X_R})}\big].
\end{equation}

Plugging (\ref{eq1}) into (\ref{eq49}) yields
\begin{eqnarray} \label{eq50} 
\mathcal{L}_{{I_{X_R}}}(s)&=&\mathbb{E}\Bigg[{\exp\Bigg(-\sum_{x\in\Phi_{X_R}}sP\vert h_{Rx}\vert^2 l_{Rx} \Bigg)}\Bigg]\nonumber\\
&=& \mathbb{E}\Bigg[\prod_{x\in\Phi_{X_R}} \exp\Bigg(-sP\vert h_{Rx}\vert^2l_{Rx}\Bigg)\Bigg]\nonumber\\
&\overset{(a)}{=}&\mathbb{E}\Bigg[\prod_{x\in\Phi_{X_R}}\mathbb{E}_{\vert  h_{Rx}\vert^2}\Bigg\lbrace \exp\Bigg(-sP\vert h_{Rx}\vert^2l_{Rx}\Bigg)\Bigg\rbrace\Bigg]\nonumber\\
&\overset{(b)}{=}&\mathbb{E}\Bigg[\prod_{x\in\Phi_{X_R}}\dfrac{p}{1+sP l_{Rx}}+1-p\Bigg]\nonumber
\end{eqnarray}
\begin{eqnarray} 
&\overset{(c)}{=}&\exp\Bigg(-\lambda_{X}\displaystyle\int_{\mathcal{B}}\Bigg[1-\bigg(\dfrac{p}{1+sPl_{Rx}}+1-p\bigg)\Bigg]dx\Bigg)\nonumber\\
&=&\exp\Bigg(-p\lambda_{X}\displaystyle\int_{\mathcal{B}}\dfrac{1}{1+1/sPl_{Rx}}dx\Bigg), \nonumber
\end{eqnarray}
%\begin{eqnarray}
%\mathcal{L}_{{I_{X_R}}}(s)=\mathbb{E}\Bigg[{\exp\Bigg(-\sum_{x\in\Phi_{X_R}}sP\vert h_{Rx}\vert^2 l_{Rx}\mathds{1}\Big\lbrace x\in \Phi_{X_{R}} \Big\rbrace  \Bigg)}\Bigg]\nonumber\\
%= \mathbb{E}\Bigg[\prod_{x\in\Phi_{X_R}} \exp\Bigg(-sP\vert h_{Rx}\vert^2l_{Rx}\mathds{1}\Big\lbrace x\in \Phi_{X_{R}}\Big\rbrace\Bigg)\Bigg]\nonumber\\
%\overset{(a)}{=}\mathbb{E}\Bigg[\prod_{x\in\Phi_{X_R}}\mathbb{E}_{\vert  h\vert^2}\Bigg\lbrace \exp\Bigg(-sP\vert h_{Rx}\vert^2l_{Rx}\mathds{1}\Big\lbrace x\in \Phi_{X_{R}}\Big\rbrace\Bigg)\Bigg\rbrace\Bigg]\nonumber\\
%\overset{(b)}{=}\mathbb{E}\Bigg[\prod_{x\in\Phi_{X_R}}\dfrac{p}{1+sP l_{Rx}}+1-p\Bigg]\nonumber\\
%\overset{(c)}{=}\exp\Bigg(-\lambda_{X}\displaystyle\int_{\mathcal{B}}\Bigg[1-\bigg(\dfrac{p}{1+sPl_{Rx}}+1-p\bigg)\Bigg]dx\Bigg)\nonumber\\
%=\exp\Bigg(-p\lambda_{X}\displaystyle\int_{\mathcal{B}}\dfrac{1}{1+1/sPl_{Rx}}dx\Bigg), \nonumber
%\end{eqnarray}

where (a) follows from having independent fading; (b) follows from calculating the expectation over $|h_{Rx}|^2$ which follows an exponential distribution with unit mean, and then calculating the expectation over the indicator function $\mathds{1}$; (c) follows from the probability generating functional (PGFL) of a PPP \cite{haenggi2012stochastic}.
Then (\ref{eq48}) can be expressed as
\begin{equation} \label{eq51} 
\mathbb{E}_{I_X}\big[ e^{-s I_{X_R}}\big]\mathbb{E}_{I_X}\big[ e^{-b I_{X_D}}\big]=\mathcal{L}_{I_{X_R}}(s)\mathcal{L}_{I_{X_D}}(b),
\end{equation}
where:
\begin{equation}\label{eq51} 
\mathcal{L}_{I_{X_R}}(s)=\exp\Bigg(-\emph{p}\lambda_{X}\int_\mathcal{B}\dfrac{1}{1+\big(A\Vert \textit{x}-R \Vert^\alpha\big)/sP}dx\Bigg)
\end{equation}
\begin{equation}\label{eq52} 
\mathcal{L}_{I_{X_D}}(s)=\exp\Bigg(-\emph{p}\lambda_{X}\int_\mathcal{B}\dfrac{1}{1+\big(A\Vert \textit{x}-D \Vert^\alpha\big)/sP}dx\Bigg).
\end{equation}
In the same way, when the interference originating from the Y road at the relay and the destination are generated from two independent sets, the following expectation can be written as
\begin{equation} \label{eq53} 
\mathbb{E}_{I_Y}\big[ e^{-s I_{Y_R}}\big]\mathbb{E}_{I_Y}\big[ e^{-b I_{Y_D}}\big]=\mathcal{L}_{I_{Y_R}}(s)\mathcal{L}_{I_{Y_D}}(b),
\end{equation}
where
\begin{equation}\label{eq54} 
\mathcal{L}_{I_{Y_R}}(s)=\exp\Bigg(-\emph{p}\lambda_{Y}\int_\mathcal{B}\dfrac{1}{1+\big(A\Vert \textit{y}-R \Vert^\alpha\big)/sP}dy\Bigg)
\end{equation}
\begin{equation}\label{eq55} 
\mathcal{L}_{I_{Y_D}}(s)=\exp\Bigg(-\emph{p}\lambda_{Y}\int_\mathcal{B}\dfrac{1}{1+\big(A\Vert \textit{y}-D \Vert^\alpha\big)/sP}dy\Bigg).
\end{equation}
After substituting all the expressions of the expectation in (\ref{eq17}), (\ref{eq18}) and (\ref{eq19}), we obtain (\ref{eq20}), (\ref{eq21}) and (\ref{eq22}).
\section{}\label{appE}
\begin{center}
PROOF OF THEOREM 2
\end{center}
When the interference at the relay and the destination are generated from the same set, the equality in (\ref{eq48}) does not hold true. Then, the expectation in left side of (\ref{eq48}) will be expressed as
\begin{eqnarray} \label{} 
\mathbb{E}_{I_X}\bigg[ e^{-\big(s I_{X_R}+b I_{X_D}\big)}\bigg]
&=&\mathbb{E}_{I_X}\Bigg[\prod_{x\in\Phi_{X}}\mathbb{E}_{\vert  h \vert^2}\Big[e^{- sP\vert h_{Rx}\vert^2l_{Rx}+bP\vert h_{Dx}\vert^2l_{Dx}}\Big]\Bigg]\nonumber \nonumber
\end{eqnarray}
\begin{eqnarray}
&=&\mathbb{E}_{I_X}
\Bigg[\prod_{x\in\Phi_{X}}\dfrac{p}{\big(1+sPl_{Rx}\big)\big(1+bPl_{Dx}\big)}+1-p\Bigg]\nonumber\\
&=&\exp\Bigg(-\lambda_{X}\int_\mathcal{B}1-\Bigg[\dfrac{p}{(1+sPl_{Rx})(1+bPl_{Dx})}+1-p\Bigg]dx\Bigg)\nonumber\\
&=&\exp\Bigg(-p\lambda_{X}\int_\mathcal{B}\dfrac{sP l_{Rx}}{1+sP l_{Rx}}+\dfrac{bP l_{Dx}}{1+bP l_{Dx}}-
\dfrac{sbP^2 l_{Rx} l_{Dx}}{\big(1+sP l_{Rx}\big)\big(1+bP l_{Dx}\big)}dx\Bigg)\nonumber
\end{eqnarray}
\begin{eqnarray} \label{eq56}
=\quad\exp\Bigg(-p\lambda_{X}\int_\mathcal{B}\dfrac{dx}{1+1/sP l_{Rx}}\Bigg)
\exp\Bigg(-p\lambda_{X}\int_\mathcal{B}\dfrac{dx}{1+1/bP l_{Dx}}\Bigg)\nonumber\\\exp\Bigg(p\lambda_{X}\int_\mathcal{B}\dfrac{sbP^2 l_{Rx} l_{Dx}}{\big(1+sP l_{Rx}\big)\big(1+bP l_{Dx}\big)}dx\Bigg).
\end{eqnarray}
Then (\ref{eq56}) can be written as
\begin{equation} \label{eq57} 
\mathbb{E}_{I_X}\bigg[ e^{-\big(s I_{X_R}+b I_{X_D}\big)}\bigg]=\mathcal{L}_{I_{X_R}}(s)\mathcal{L}_{I_{X_D}}(b)\rho_{X}(s,b)= \mathcal{L}_{I_{X_R},I_{X_D}}(s,b).
\end{equation}

Following the same steps, we obtain 
\begin{equation} \label{eq58} 
\mathbb{E}_{I_Y}\bigg[ e^{-\big(s I_{Y_R}+b I_{Y_D}\big)}\bigg]=\mathcal{L}_{I_{Y_R}}(s)\mathcal{L}_{I_{Y_D}}(b)\rho_{Y}(s,b)= \mathcal{L}_{I_{Y_R},I_{Y_D}}(s,b).
\end{equation}

After substituting all the expressions of the expectation in (\ref{eq17}), (\ref{eq18}) and (\ref{eq19}), we obtain (\ref{eq233}), (\ref{eq20}), and (\ref{eq25}).

\section{}\label{appF}
\begin{center}
PROOF OF PROPOSITION 1
\end{center}

In order to calculate the Laplace transform of interference originated from the X road at the relay, we have to calculate the integral in (\ref{eq27}). We calculate the integral in (\ref{eq27}) for $\mathcal{B}=\mathbb{R}$ and $\alpha=2$. Let us take $k=sP/A^2$, $n_x=n \cos(\theta_N)$ and $n_y=n \sin(\theta_N$), then (\ref{eq27}) becomes
\begin{equation}\label{eq59} 
\mathcal{L}_{I_{X_N}}(s)=\exp\Bigg(-\emph{p}\lambda_{X}k\int_\mathbb{R}\dfrac{1}{k+n_{y}^2+(x-n_{x})^2}dx\Bigg),
\end{equation}
and the integral inside the exponential in (\ref{eq59}) equals
\begin{equation} \label{eq60} 
\int_\mathbb{R}\dfrac{1}{k+n_{y}^2+(x-n_{x})^2}dx=\dfrac{\pi}{\sqrt{n_{y}^2+k}} .
\end{equation}
Then, plugging (\ref{eq60}) into (\ref{eq59}) we obtain
\begin{equation} \label{eq61} 
\mathcal{L}_{I_{X_N}}(s)=\exp\Bigg(-\emph{p}\lambda_{X}k\dfrac{\pi}{\sqrt{n_y^2+k}}\Bigg).
\end{equation}
Finally, substituting $k$  and $n_y$ in (\ref{eq61}) yields (\ref{eq31}). Following the same steps above, and without details for the derivation, we obtain (\ref{eq32}).
\section{}\label{appG}
\begin{center}
PROOF OF PROPOSITION 2
\end{center}
We calculate the integral in (\ref{eq27}) for $\mathcal{B}=[-Z,Z]$ and $\alpha=2$. We use the same change of variables as in \textbf{Proposition 1}, then we get 
\begin{eqnarray} \label{eq62} 
\int_{-Z}^{+Z}\dfrac{1}{1+\big(A\Vert \textit{x}-N \Vert^2\big)/sP}dx=\int_{-Z}^{+Z}\dfrac{1}{1+n_{y}^{2}+(x-n_x)^{2}/k}dx\nonumber \\ =\int_{-Z}^{+Z}1-\dfrac{n_{y}^{2}+(x-n_x)^{2}/k}{1+n_{y}^{2}+(x-n_x)^{2}/k}dx=2Z-\int_{-Z}^{+Z}\dfrac{n_{y}^{2}+(x-n_x)^{2}}{k+n_{y}^{2}+(x-n_x)^{2}}dx.
\end{eqnarray}
The integral in the last equality in (\ref{eq62}) equals
\begin{multline}\label{eq63} 
\int_{-Z}^{+Z}\dfrac{n_{y}^{2}+(x-n_x)^{2}}{k+n_{y}^{2}+(x-n_x)^{2}}dx=\dfrac{2Z\sqrt{n_y^{2}+k}}{\sqrt{n_y^{2}+k}}\\
-\frac{k\arctan\Bigg( \dfrac{Z+n_x}{\sqrt{n_y^{2}+k}}\Bigg)-k\arctan\Bigg( \dfrac{Z-n_x}{\sqrt{n_y^{2}+k}}\Bigg)}{\sqrt{n_y^{2}+k}},
\end{multline}
then plugging (\ref{eq63}) into (\ref{eq62}) yields (\ref{eq33}). Following the same steps above, we obtain (\ref{eq34}).
 
%\paragraph{Functionality} The Elsevier article class is based on the standard article class and supports almost all of the functionality of that class. In addition, it features commands and options to format the
%\begin{itemize}
%\item document style
%\item baselineskip
%\item front matter
%\item keywords and MSC codes
%\item theorems, definitions and proofs
%\item lables of enumerations
%\item citation style and labeling.
%\end{itemize}
%
%\section{Front matter}
%
%The author names and affiliations could be formatted in two ways:
%\begin{enumerate}[(1)]
%\item Group the authors per affiliation.
%\item Use footnotes to indicate the affiliations.
%\end{enumerate}
%See the front matter of this document for examples. You are recommended to conform your choice to the journal you are submitting to.
%
%\section{Bibliography styles}
%
%There are various bibliography styles available. You can select the style of your choice in the preamble of this document. These styles are Elsevier styles based on standard styles like Harvard and Vancouver. Please use Bib\TeX\ to generate your bibliography and include DOIs whenever available.
%
%Here are two sample references: \cite{Feynman1963118,Dirac1953888}.

%\bibliography{mybibfile}
\bibliographystyle{ieeetr}
\bibliography{biblio}

\begin{thebibliography}{10}

\bibitem{traficsafety}
{U.S. Dept. of Transportation, National Highway Traffic Safety Administration},
  ``Traffic safety facts 2015,'' Jan. 2017.

\bibitem{sam2016vehicle}
D.~Sam, C.~Velanganni, and T.~E. Evangelin, ``A vehicle control system using a
  time synchronized hybrid vanet to reduce road accidents caused by human
  error,'' {\em Vehicular communications}, vol.~6, pp.~17--28, 2016.

\bibitem{singh2019multipath}
P.~K. Singh, S.~Sharma, S.~K. Nandi, and S.~Nandi, ``Multipath tcp for v2i
  communication in sdn controlled small cell deployment of smart city,'' {\em
  Vehicular communications}, vol.~15, pp.~1--15, 2019.

\bibitem{boquet2018adaptive}
G.~Boquet, I.~Pisa, J.~L. Vicario, A.~Morell, and J.~Serrano, ``Adaptive
  beaconing for rsu-based intersection assistance systems: Protocols analysis
  and enhancement,'' {\em Vehicular communications}, vol.~14, pp.~1--14, 2018.

\bibitem{haenggi2009interference}
M.~Haenggi, R.~K. Ganti, {\em et~al.}, ``Interference in large wireless
  networks,'' {\em Foundations and Trends{\textregistered} in Networking},
  vol.~3, no.~2, pp.~127--248, 2009.

\bibitem{ganti2009spatial}
R.~K. Ganti and M.~Haenggi, ``Spatial and temporal correlation of the
  interference in aloha ad hoc networks,'' {\em IEEE Communications Letters},
  vol.~13, no.~9, 2009.

\bibitem{tanbourgi2013cooperative}
R.~Tanbourgi, H.~J{\"a}kel, and F.~K. Jondral, ``Cooperative relaying in a
  poisson field of interferers: A diversity order analysis,'' in {\em
  Information Theory Proceedings (ISIT), 2013 IEEE International Symposium on},
  pp.~3100--3104, IEEE, 2013.

\bibitem{haenggi2009outage}
M.~Haenggi, ``Outage, local throughput, and capacity of random wireless
  networks,'' {\em IEEE Transactions on Wireless Communications}, vol.~8,
  no.~8, 2009.

\bibitem{schilcher2012temporal}
U.~Schilcher, C.~Bettstetter, and G.~Brandner, ``Temporal correlation of
  interference in wireless networks with rayleigh block fading,'' {\em IEEE
  Transactions on Mobile Computing}, vol.~11, no.~12, pp.~2109--2120, 2012.

\bibitem{tanbourgi2014effect}
R.~Tanbourgi, H.~S. Dhillon, J.~G. Andrews, and F.~K. Jondral, ``Effect of
  spatial interference correlation on the performance of maximum ratio
  combining,'' {\em IEEE Transactions on Wireless Communications}, vol.~13,
  no.~6, pp.~3307--3316, 2014.

\bibitem{tanbourgi2014dual}
R.~Tanbourgi, H.~S. Dhillon, J.~G. Andrews, and F.~K. Jondral, ``Dual-branch
  mrc receivers under spatial interference correlation and nakagami fading,''
  {\em IEEE Transactions on Communications}, vol.~62, no.~6, pp.~1830--1844,
  2014.

\bibitem{ikki2013regenerative}
S.~S. Ikki, P.~Ubaidulla, and S.~A{\"\i}ssa, ``Regenerative cooperative
  diversity networks with co-channel interference: Performance analysis and
  optimal energy allocation,'' {\em IEEE Transactions on Vehicular Technology},
  vol.~62, no.~2, pp.~896--902, 2013.

\bibitem{altieri2014outage}
A.~Altieri, L.~R. Vega, P.~Piantanida, and C.~G. Galarza, ``On the outage
  probability of the full-duplex interference-limited relay channel,'' {\em
  arXiv preprint arXiv:1403.7317}, 2014.

\bibitem{schilcher2013does}
U.~Schilcher, S.~Toumpis, A.~Crismani, G.~Brandner, and C.~Bettstetter, ``How
  does interference dynamics influence packet delivery in cooperative
  relaying?,'' in {\em Proceedings of the 16th ACM international conference on
  Modeling, analysis \& simulation of wireless and mobile systems},
  pp.~347--354, ACM, 2013.

\bibitem{crismani2015cooperative}
A.~Crismani, S.~Toumpis, U.~Schilcher, G.~Brandner, and C.~Bettstetter,
  ``Cooperative relaying under spatially and temporally correlated
  interference,'' {\em IEEE Transactions on Vehicular Technology}, vol.~64,
  no.~10, pp.~4655--4669, 2015.

\bibitem{blaszczyszyn2009performance}
B.~Blaszczyszyn, P.~Muhlethaler, and Y.~Toor, ``Performance of mac protocols in
  linear vanets under different attenuation and fading conditions,'' in {\em
  Intelligent Transportation Systems, 2009. ITSC'09. 12th International IEEE
  Conference on}, pp.~1--6, IEEE, 2009.

\bibitem{blaszczyszyn2013stochastic}
B.~B{\l}aszczyszyn, P.~M{\"u}hlethaler, and Y.~Toor, ``Stochastic analysis of
  aloha in vehicular ad hoc networks,'' {\em Annals of
  telecommunications-Annales des t{\'e}l{\'e}communications}, vol.~68, no.~1-2,
  pp.~95--106, 2013.

\bibitem{blaszczyszyn2012vehicular}
B.~Blaszczyszyn, P.~Muhlethaler, and N.~Achir, ``Vehicular ad-hoc networks
  using slotted aloha: point-to-point, emergency and broadcast
  communications,'' in {\em Wireless Days (WD), 2012 IFIP}, pp.~1--6, IEEE,
  2012.

\bibitem{farooq2016stochastic}
M.~J. Farooq, H.~ElSawy, and M.-S. Alouini, ``A stochastic geometry model for
  multi-hop highway vehicular communication,'' {\em IEEE Transactions on
  Wireless Communications}, vol.~15, no.~3, pp.~2276--2291, 2016.

\bibitem{jiang2016information}
C.~Jiang, H.~Zhang, Z.~Han, Y.~Ren, V.~C. Leung, and L.~Hanzo,
  ``Information-sharing outage-probability analysis of vehicular networks,''
  {\em IEEE Transactions on Vehicular Technology}, vol.~65, no.~12,
  pp.~9479--9492, 2016.

\bibitem{tassi2017modeling}
A.~Tassi, M.~Egan, R.~J. Piechocki, and A.~Nix, ``Modeling and design of
  millimeter-wave networks for highway vehicular communication,'' {\em IEEE
  Transactions on Vehicular Technology}, vol.~66, no.~12, pp.~10676--10691,
  2017.

\bibitem{steinmetz2015stochastic}
E.~Steinmetz, M.~Wildemeersch, T.~Q. Quek, and H.~Wymeersch, ``A stochastic
  geometry model for vehicular communication near intersections,'' in {\em
  Globecom Workshops (GC Wkshps), 2015 IEEE}, pp.~1--6, IEEE, 2015.

\bibitem{abdulla2016vehicle}
M.~Abdulla, E.~Steinmetz, and H.~Wymeersch, ``Vehicle-to-vehicle communications
  with urban intersection path loss models,'' in {\em Globecom Workshops (GC
  Wkshps), 2016 IEEE}, pp.~1--6, IEEE, 2016.

\bibitem{abdulla2017fine}
M.~Abdulla and H.~Wymeersch, ``Fine-grained reliability for v2v communications
  around suburban and urban intersections,'' {\em arXiv preprint
  arXiv:1706.10011}, 2017.

\bibitem{jeyaraj2017reliability}
J.~P. Jeyaraj and M.~Haenggi, ``Reliability analysis of v2v communications on
  orthogonal street systems,'' in {\em GLOBECOM 2017-2017 IEEE Global
  Communications Conference}, pp.~1--6, IEEE, 2017.

\bibitem{kimura2017theoretical}
T.~Kimura and H.~Saito, ``Theoretical interference analysis of inter-vehicular
  communication at intersection with power control,'' {\em Computer
  Communications}, 2017.

\bibitem{article}
B.~E.~Y. Belmekki, A.~Hamza, and B.~Escrig, ``Cooperative vehicular
  communications at intersections over nakagami-m fading channels,'' {\em
  Vehicular Communications}, p.~doi:10.1016/j.vehcom.2019.100165, 07 2019.

\bibitem{WCNC}
B.~E.~Y. Belmekki, A.~Hamza, and B.~Escrig, ``Outage performance of {NOMA} at
  road intersections using stochastic geometry,'' in {\em 2019 IEEE Wireless
  Communications and Networking Conference (WCNC) (IEEE WCNC 2019)}, pp.~1--6,
  IEEE, 2019.

\bibitem{J3}
B.~E.~Y. Belmekki, A.~Hamza, and B.~Escrig, ``On the performance of 5g
  non-orthogonal multiple access for vehicular communications at road
  intersections,'' {\em Vehicular Communications},
  p.~doi:10.1016/j.vehcom.2019.100202, 2019.

\bibitem{VTC}
B.~E.~Y. Belmekki, A.~Hamza, and B.~Escrig, ``On the outage probability of
  cooperative 5g noma at intersections,'' in {\em 2019 IEEE 89th Vehicular
  Technology Conference (VTC2019-Spring)}, pp.~1--6, IEEE, 2019.

\bibitem{J4}
B.~E.~Y. Belmekki, A.~Hamza, and B.~Escrig, ``Performance analysis of
  cooperative noma at intersections for vehicular communications in the
  presence of interference,'' {\em Ad hoc Networks},
  p.~doi:10.1016/j.adhoc.2019.102036, 2019.

\bibitem{WiMob}
B.~E.~Y. Belmekki, A.~Hamza, and B.~Escrig, ``Outage analysis of cooperative
  noma using maximum ratio combining at intersections,'' in {\em IEEE 15th Int.
  Conf. Wireless Mobile Comput. Netw. Commun. (WiMob)}, pp.~1--6, IEEE, 2019.

\bibitem{NoMa}
B.~E.~Y. Belmekki, A.~Hamza, and B.~Escrig, ``Non-orthogonal multiple access
  performance for millimeter wave in vehicular communications,'' {\em arXiv
  preprint arXiv:1909.12392}, 2019.

\bibitem{NoMa2}
B.~E.~Y. Belmekki, A.~Hamza, and B.~Escrig, ``Outage analysis of cooperative
  noma in millimeter wave vehicular network at intersections,'' {\em arXiv
  preprint arXiv:1904.11022}, 2019.

\bibitem{cheng2007mobile}
L.~Cheng, B.~E. Henty, D.~D. Stancil, F.~Bai, and P.~Mudalige, ``Mobile
  vehicle-to-vehicle narrow-band channel measurement and characterization of
  the 5.9 ghz dedicated short range communication (dsrc) frequency band,'' {\em
  IEEE Journal on Selected Areas in Communications}, vol.~25, no.~8,
  pp.~1501--1516, 2007.

\bibitem{haenggi2012diversity}
M.~Haenggi, ``Diversity loss due to interference correlation,'' {\em IEEE
  Communications Letters}, vol.~16, no.~10, pp.~1600--1603, 2012.

\bibitem{rappaport2002wireless}
T.~S. Rappaport, ``Wireless communications--principles and practice, (the book
  end),'' {\em Microwave Journal}, vol.~45, no.~12, pp.~128--129, 2002.

\bibitem{haenggi2012stochastic}
M.~Haenggi, {\em Stochastic geometry for wireless networks}.
\newblock Cambridge University Press, 2012.

\end{thebibliography}
\end{document}